\documentclass{amsart}

\pdfoutput=1

\usepackage{amsmath,amssymb}
\usepackage{graphicx}
\usepackage{subfigure}
\usepackage{url}
\usepackage{listings}
\usepackage{multirow}
\usepackage{caption}
\usepackage{color}
\usepackage{hyperref}

\begin{document}

\hypersetup{
  pdfauthor={S.S. Craciunas, C.M. Kirsch, H. Payer, H. R{\"o}ck, A. Sokolova},
  pdftitle={Concurrency and Scalability versus Fragmentation and Compaction with Compact-fit},
  pdfsubject={Computer Science, Memory Management},
  pdfcreator={C.M. Kirsch}
}

\title[Concurrency and Scalability with Compact-fit]{Concurrency and Scalability versus Fragmentation and Compaction with Compact-fit}

\author[S.S. Craciunas]{Silviu S. Craciunas}
\author[C.M. Kirsch]{Christoph M. Kirsch}
\author[H. Payer]{Hannes Payer}
\author[H. R{\"o}ck]{Harald R\"ock}
\author[A. Sokolova]{Ana Sokolova}

\address{Department of Computer Sciences\\ University of Salzburg\\ Austria}
\email{ck@cs.uni-salzburg.at}
\urladdr{cs.uni-salzburg.at/~ck}

\begin{abstract}
We study, formally and experimentally, the trade-off in temporal and
spatial overhead when managing contiguous blocks of memory using the
explicit, dynamic and real-time heap management system Compact-fit
(CF).  The key property of CF is that temporal and spatial overhead
can be bounded, related, and predicted in constant time through the
notion of partial and incremental compaction.  Partial compaction
determines the maximally tolerated degree of memory fragmentation.
Incremental compaction of objects, introduced here, determines the
maximal amount of memory involved in any, logically atomic, portion of
a compaction operation.  We explore CF's potential application space
on (1)~multiprocessor and multicore systems as well as on
(2)~memory-constrained uniprocessor systems.  For~(1), we argue that
little or no compaction is likely to avoid the worst case in temporal
as well as spatial overhead but also observe that scalability only
improves by a constant factor.  Scalability can be further improved
significantly by reducing overall data sharing through separate
instances of Compact-fit.  For~(2), we observe that incremental
compaction can effectively trade-off throughput and memory
fragmentation for lower latency.
\end{abstract}

\maketitle


\section{Introduction}
\label{sec:intro}

Compact-fit (CF)~\cite{USENIX08} is an explicit, dynamic, and real-time
heap management system also known as a memory allocator.  Heap
management solves the problem of allocating and deallocating memory
objects of possibly different size where the order in which the
objects are allocated and deallocated may be arbitrary.  It is dynamic
if allocation and deallocation happens at runtime, as opposed to
static, so-called pre-allocation, which may only be done if the
amount of memory needed for program execution can be bounded at
compile time.  It is real-time if the time to allocate, deallocate,
and access a memory object is either constant or at most proportional
to the size of the object, independent of the overall state of memory
and in particular the order in which objects are allocated and
deallocated.

Heap management is explicit if deallocation must be invoked by the
program using the system.  The use of explicit heap management may
therefore suffer from the well-known phenomena of memory leaks, where
memory objects are continuously allocated but never deallocated, and
so-called dangling pointers to memory objects that have been
deallocated prematurely and may lead to undefined program behavior
when accessed.  Compact-fit is not an exception.  It does not address
the problems of memory leaks and dangling pointers.

Implicit heap management solves the problem of dangling pointers by
first determining when to deallocate memory objects safely and then
using an underlying explicit heap management to actually deallocate
these objects.  Garbage collectors are implicit heap management
systems which logically operate in two phases that are performed
repeatedly.  In the first phase allocated but unreachable memory
objects are determined, either directly through reference counting and
cycle detection, or indirectly through tracing, or some combination of
both~\cite{Bac4}.  A memory object is unreachable if the program has
no means of accessing the object, neither directly through some
reference nor transitively through other, reachable memory objects.
An unreachable memory object may thus be deallocated safely without
introducing dangling pointers.

In the second phase unreachable memory objects are deallocated which
is done by an underlying memory allocator.  In other words, a garbage
collector implicitly uses a memory allocator which is, however, often
tightly integrated with the garbage collector for performance reasons.
Compact-fit is closely related to the integrated memory allocator of
the real-time garbage collector Metronome~\cite{Bac2}.

Garbage collectors solve the problem of dangling pointers but not the
problem of memory leaks: a program may continuously allocate memory
objects to which it maintains references, e.g. by inserting the objects
into an ever-growing hashtable and never removing them again.  The
result is a so-called reachable memory leak which garbage collectors
cannot avoid.  In other words, using implicit heap management does not
free a program from deallocating memory objects.  It only makes
deallocation implicit (remove references and return from procedure
calls rather than deallocate explicitly) and is therefore safe, i.e.,
the program still needs to go through the otherwise ever-growing
hashtable and remove obsolete data from time to time.  Fundamentally,
reachable memory leaks are not detected by garbage collectors because
reachability is only an overapproximation of memory liveness which
itself is undecidable: after some time a memory object may never be
accessed again but still remain reachable.

The true power of garbage collection is that it makes the use of the
heap compositional.  Programs may allocate memory objects, even in
imported library code, and pass references to them around without
keeping track of when to deallocate the objects as long as the
references are eventually removed when the objects are not needed
anymore.  Compositionality of the heap is key to large-scale program
design and particularly useful in concurrent programs where keeping
track of when shared memory may be deallocated is especially
difficult.

The price to pay for garbage collection is temporal and spatial
overhead: computing unreachability (directly or indirectly) is
proportional to the size of live, i.e., reachable memory (in the
presence of cyclic references which is the case in most non-trivial
applications), resulting in lagged deallocation of unreachable memory
and thus increased memory consumption.  Temporal overhead, when
created in so-called stop-the-world fashion, precludes real-time
applications.  Spatial overhead precludes embedded applications, in
particular if deallocation not only lags unreachability but also
results in uncontrolled memory fragmentation which may also occur in
explicit heap management without any garbage collection.

Temporal and spatial overhead of dynamic heap management, explicit or
implicit, cannot be avoided but it can be bounded!  The key to
enabling dynamic heap management in real-time and embedded
applications is to make it incremental and to bound memory
fragmentation.  Heap management is incremental if it may be done in
phases whose durations are constant and which may be interleaved with
program execution.  The maximum duration of a heap management phase
determines the latency introduced by heap management and thus directly
defines the compatible class of real-time applications.  The drawback
of incremental heap management is lower throughput since the sum of
the phases of a heap management operation is generally larger than the
duration of the operation when not interrupted.  Note that the focus
of this paper is on making explicit heap management incremental while
bounding memory fragmentation, which is one of the two fundamental
prerequisites for incremental garbage collection.  The other
prerequisite is incrementally computing unreachability which is
addressed elsewhere, e.g. in Metronome~\cite{Bac2}.

Memory fragmentation is the phenomenon of unoccupied memory blocks
being dispersed in memory (external fragmentation) and/or designated
through partitioning (internal fragmentation).  If contiguous memory
blocks of different size may be allocated and deallocated in arbitrary
order, uncontrolled memory fragmentation may lead to unbounded gross
memory consumption even if net memory consumption is bounded.

Compact-fit avoids external fragmentation and bounds internal
fragmentation through partitioning and so-called partial compaction.
Upon deallocating a memory object partial compaction may move another
same-size object into its place but only if a given threshold on
fragmentation is exceeded.  In this case, deallocation takes time
linear in the size of the deallocated object.  Otherwise, deallocation
is constant-time.  Memory allocation as well as access are always
constant-time.  The principle topic of this paper is to make partial
compaction incremental such that objects are moved incrementally in
phases of constant duration and yet may still be accessed in constant
time in between compaction phases.  The result is what we call
incremental Compact-fit, the first memory allocator that bounds the
full spectrum of temporal and spatial overhead of memory allocation,
deallocation, and access in terms of configurable constants.  With
incremental Compact-fit the duration of any heap management activity
as well as the degree of memory fragmentation are bounded by
constants, which makes this allocator the principle choice for any
application in which constant bounds on both temporal and spatial
overhead are required.

Note that there are memory allocators that either bound temporal
overhead in terms of constants such as Half-fit~\cite{Oga1} and
TLSF~\cite{Mas1} or else spatial overhead such as the allocator of the
Jamaica VM~\cite{Sie2} but not both. Half-fit and TLSF provide
constant-time memory allocation, deallocation, and access, but only
control and not bound memory fragmentation through coalescing
neighboring, unoccupied memory blocks.  The Jamaica allocator avoids
external fragmentation and bounds internal fragmentation but at the
expense of constant-time memory allocation, deallocation, and access
as well as memory locality by allocating small, same-size but
generally dispersed memory blocks and assembling them into larger
memory objects through trees (logarithmic-time access) or lists
(linear-time access) that fit the requested size.

Next, we discuss the design principles and features of Compact-fit
before providing an overview of the rest of the paper.

\subsection{Compact-fit}

Compact-fit partitions memory into virtual pages of equal size by
maintaining a list of free pages and a segregated list of finitely
many so-called size-classes where each size-class is a doubly-linked
list of used pages that are further partitioned into virtual,
so-called page-blocks of equal and unique size.  A memory object is
allocated as contiguous block of memory in a free page-block of the
size-class with the smallest page-block size that still fits the
object.  Memory allocation, deallocation, and access takes constant
time (unless compaction is necessary when deallocating, which takes
linear time in the size of the deallocated object).  Allocation of
memory objects larger than the page size is not part of CF itself but
may be done on top of CF by array, tree-, or list-based data
structures that combine sufficiently many pages to accommodate large
objects resulting in allocation and deallocation times that are linear
and memory access times that are constant, logarithmic, or linear,
respectively, in the size of the objects.  However, we do not consider
large-object management here.

The size-class concept is generally subject to fragmentation through
partitioning, that is, to bounded page-block-internal, page-internal,
and size-external fragmentation~\cite{Bac1}, but enables CF to keep
memory size-class-compact at all times~\cite{USENIX08}.  Memory is
size-class-compact if each of its size-classes is compact.  A
size-class is compact with respect to a so-called partial compaction
bound $\kappa$ if the size-class contains only non-empty pages of
which at most $\kappa$ are not-full.  A size-class is said to be
totally compact, fully compact, or partially compact if it is compact
with respect to $\kappa=0$, $\kappa=1$, or $\kappa>1$, respectively.
Note that, as opposed to the leftover space caused by fragmentation
through partitioning, which is wasted for any request, the free space
in not-full pages of a size-class, called size-class fragmentation, is
wasted for any request but the requests that actually match the
size-class.  Partial compaction can only control the degree of
size-class fragmentation.

CF always keeps all size-classes compact with respect to individual,
per-size-class partial compaction bounds $\kappa>0$.  Overall memory
fragmentation is therefore bounded and predictable in constant time.
Note that $\kappa=\infty$ is also permissible and means that any
number of not-full pages in a size-class is tolerated.  A memory
object is allocated, in constant time, in a free page-block either of
a not-full page of the adequate size-class (implicitly compacting
allocation), or else, if there is no not-full page in the size-class,
of a free page that is then removed from the list of free pages and
assigned to the size-class (non-compacting allocation).  A memory
object is deallocated, either in constant time, by marking the
page-block used by the object as free, if the size-class remains
partially compact (non-compacting deallocation), or else in linear
time in the size of the object, by marking a used page-block of a
not-full, so-called source page as free after copying the content of
that (source) page-block to the (target) page-block used by the
object, which, in this case, must be located in a full, so-called
target page (compacting deallocation).  If the page in which a
page-block was marked as free becomes empty, the page is removed from
the size-class and returned to the list of free pages.

In order to facilitate compacting memory that may contain references
in time linear in the size of the moved objects, CF maintains a map
(A2C) from abstract object addresses that do not change when moving
objects, also referred to as handles, to the concrete object addresses
in memory.  Objects may only refer to other objects using their
abstract addresses, which implies that memory access requires one
level of indirection, unless compaction is turned off with
$\kappa=\infty$.  As a result, whenever an object is moved in memory,
only its concrete address in the A2C map needs to be updated.  CF
stores the abstract address of each object in the object itself so
that the object's entry in the A2C map can be determined in constant
time.  Otherwise, determining the abstract addresses of objects
selected for compaction, for which only the concrete addresses are
known, would require searching the A2C map.

There is also a non-moving version of CF~\cite{USENIX08}, which
virtualizes the concrete address space using an additional level of
indirection that merely requires reprogramming a map (V2P) from
virtual to physical addresses upon compaction instead of moving the
actual content of the objects.  Since objects do not move, their
physical addresses can be used to generate unique abstract addresses,
which avoids storing abstract addresses in objects.  Nevertheless, in
the worst case, the V2P map requires just as much memory as the object
storage for abstract addresses.  Moreover, experiments have shown that
the non-moving version of CF may only pay off when used for larger
objects~\cite{USENIX08}.  In the rest of the article, we only consider
the moving version of CF.

\subsection{Overview}

After discussing related work (Section~\ref{sec:rel}) and discussing
the previously described, moving (and non-incremental) version of CF
in detail (Section~\ref{sec:non-inc-cf}), we first argue
probabilistically that, for particular mutator behavior, both compaction and
worst-case size-class fragmentation are less likely to happen with
increasing partial compaction bounds~$\kappa$.  For systems whose
memory resources are less constrained and applications that do not
require tight guarantees, partial compaction may therefore be set to
large $\kappa$, or even turned off entirely.  This observation has
lead us to develop an optimized, non-compacting version of CF without
abstract addressing that does not maintain the A2C map and can
therefore be used in any application without modifications.
Macrobenchmarks show that the optimized version performs almost as
fast as other constant-time state-of-the-art memory
allocators. Moreover, less than 5\% of the fragmentation can be
attributed to size-class fragmentation and the rest to fragmentation
through partitioning (Section~\ref{sec:exp}).  We argue that
partitioning memory as in CF still has the benefit of being subject to
a probabilistic and not just an experimental fragmentation analysis
(Section~\ref{sec:prob}), at the expense of increased memory
consumption.

We then introduce incremental CF for slow systems, at the other end of
the spectrum, whose memory resources are constrained and that run
applications requiring tight guarantees, in particular on system
latency and memory consumption (Section~\ref{sec:inc-cf}).
Incremental CF uses a global compaction increment $\iota>0$, which
breaks up compaction into logically atomic operations that do not move
more than $\iota$ bytes at a time.  If $n$ is the degree of
concurrency, then there may be at most $n$ pending incremental
compaction operations moving objects stored in $n$ source page-blocks
from $n$ source pages to $n$ target pages.  The memory occupied by the
$n$ source page-blocks causes so-called transient size-class
fragmentation in the $n$ source pages.  The key result is that the
time complexity of memory allocation, deallocation, and access remains
asymptotically the same as with non-incremental CF while overall
memory fragmentation is still bounded and predictable in constant time
(Section~\ref{sec:complexity}).  Incremental CF may improve system
latency at the expense of allocation and deallocation throughput and
transient size-class fragmentation (Section~\ref{sec:exp}).

\begin{figure}
\begin{center}
\includegraphics[width=\textwidth]{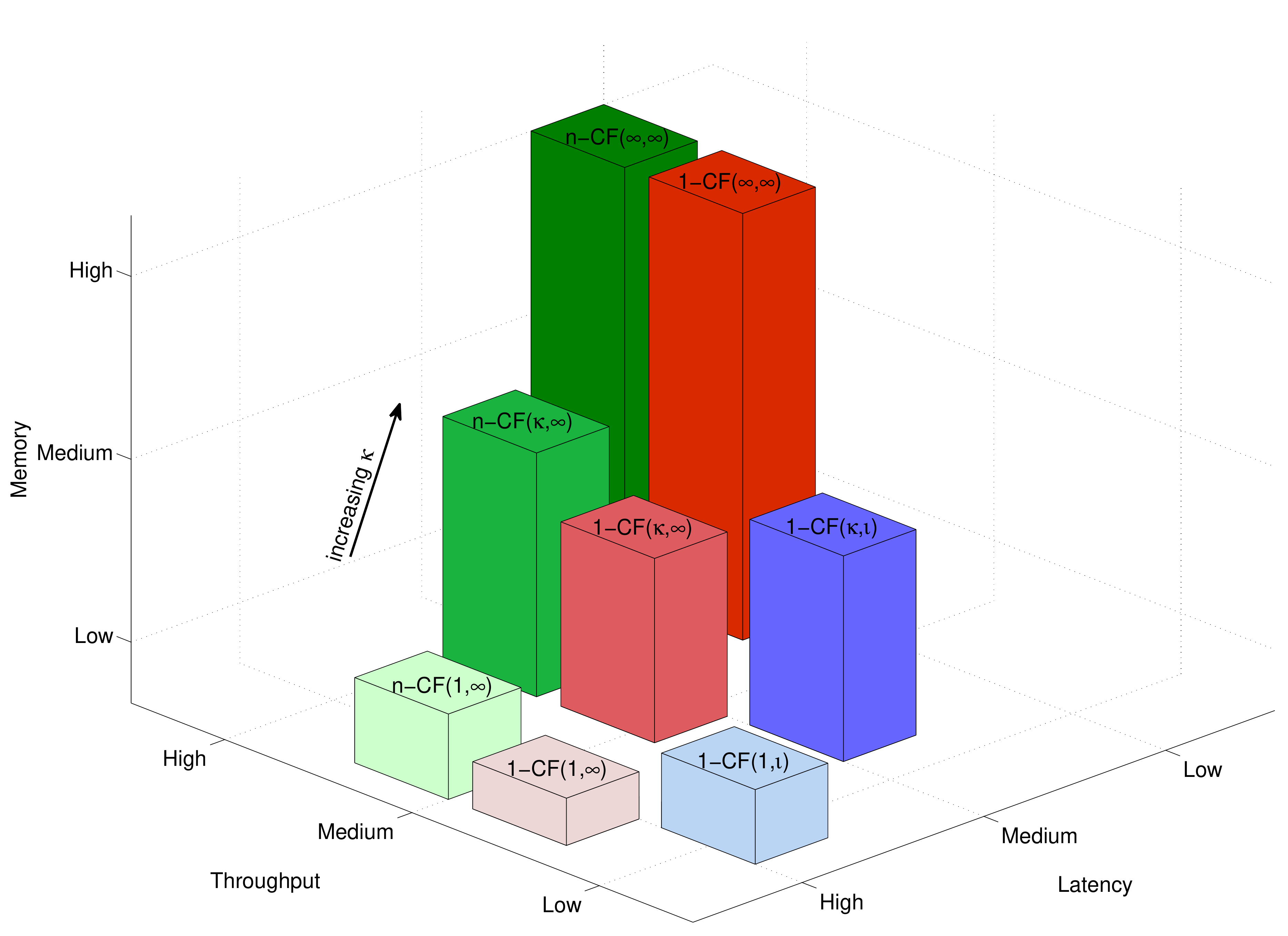}
\caption{\label{fig:tradeoff}{Allocation and} deallocation throughput,
  system latency, and memory fragmentation with different versions and
  configurations of Compact-fit}
\end{center}
\end{figure}

Figure~\ref{fig:tradeoff} gives an intuitive overview of the effect of
different versions and configurations of CF on allocation and
deallocation throughput, system latency, and memory fragmentation.\footnote{{See Section~\ref{sec:complexity} for a table with allocation/deallocation complexities of each version.}}  A
configuration $1$-CF$(\kappa,\iota)$ denotes a single instance of a CF
system with a per-size-class partial compaction bound $\kappa>0$ and a
global compaction increment~$\iota>0$.  The instance may be shared by
concurrently running threads using a number of different, standard
synchronization techniques (Section~\ref{sec:impl}).  Incremental
compaction is off if $\iota=\infty$.  Partial compaction is off if
$\kappa=\infty$, which implies that incremental compaction is also
off.  Full compaction is on if $\kappa=1$.  The fully compacting,
non-incremental $1$-CF$(1,\infty)$ configuration minimizes memory
fragmentation at the expense of throughput and latency.  In
comparison, the fully compacting, incremental $1$-CF$(1,\iota)$
configuration may require more memory because of transient size-class
fragmentation and provide less throughput but may reduce latency.
With $\kappa>1$, memory fragmentation may go up proportionally to
$\kappa$ with both configurations while throughput may be higher and
latency may be lower as there may be fewer compaction operations.  The
non-compacting $1$-CF$(\infty,\infty)$ configuration may provide even
higher throughput and lower latency but may also consume even more
memory.  The key advantage of this configuration is that it may be
optimized as mentioned above.

A configuration $n$-CF$(\kappa,\iota)$ denotes $n$ instances of a CF
system, one for each of $n$ threads, which is meant to improve
scalability on multiprocessor and multicore systems
(Section~\ref{sec:impl}).  Compared to the single-instance
configurations, throughput may be higher but memory fragmentation may
also go up with the compacting configurations since partial compaction
bounds are enforced per instance and therefore per thread.  Our
experiments show that partial compaction on fast systems may only have
an effect on scalability by a constant factor since the time required
to perform a single compaction operation on such systems is close to
the time required to perform any other CF operation, independently of
the size of the involved object.  More relevant to scalability is the
degree of data sharing, in particular, through the A2C map
(Section~\ref{sec:exp}).

The contributions of this article are the design, implementation, and
comprehensive, formal and experimental evaluation of concurrent
versions of (1)~an optimized, non-compacting CF system, (2)~the
previously described, compacting, non-incremental CF
system~\cite{USENIX08}, and (3)~a new, compacting, incremental CF
system.

\section{Related Work}
\label{sec:rel}




Scalability of concurrent memory allocators~\cite{Ber1} and garbage collection
systems~\cite{Gidra11,Payer12a} is {the} key for high performance in parallel
environments.  We relate our work to dynamic heap management systems of
different kinds: explicit sequential allocators, explicit concurrent
allocators, and concurrent garbage-collection-based systems with compaction
(cf.~\cite{Jon1} for an extensive online bibliography).

Most of the established explicit sequential dynamic heap management
systems~\cite{Mas2,Pua1} are optimized to offer excellent best-case
and average-case response times, but in the worst-case are unbounded
in the size of the memory allocation or deallocation request, i.e.,
depend on the global state of memory.  The best known are First-fit,
Best-fit~\cite{Knu1} and DL~\cite{Dou1} with allocation times
depending on the global state of memory.  Half-fit~\cite{Oga1} and
TLSF~\cite{Mas1} are exceptions offering constant response-time bounds
for allocation and deallocation, but even they may suffer from
unbounded and unpredictable memory fragmentation.

Several concurrent dynamic memory allocators have been designed for
scalable performance on multiprocessor systems. Hoard~\cite{Ber1}
provides fast and scalable memory allocation and deallocation
operations, using locks for synchronization and avoiding false sharing
of cache lines. A lock-free memory allocator with lower latency based
on the principles of Hoard is given in~\cite{Mic1}. A partly lock-free
non-portable memory allocator, which requires special operating system
support, is discussed in~\cite{Dic1}. McRT-Malloc~\cite{Hud2} is a
non-blocking scalable heap management algorithm, which avoids atomic
operations on typical code paths by accessing only thread-local data
and uses the same memory layout (pages and size-classes) as CF. None
of these systems provides temporal or spatial guarantees.

Incremental compaction typically performs the compaction phase of a garbage
collection cycle incrementally, i.e., multiple objects are moved atomically.
The incremental compaction algorithms discussed in this paragraph and the
following paragraph are based on that concept. Our incremental compaction
approach in CF is different. CF allows to move a single object incrementally,
which may reduce the latency of a compaction operation even further.  There are
many concurrent compaction strategies implemented in garbage-collected systems,
which do not provide temporal or spatial guarantees.  In~\cite{Flo1} a parallel
stop-the-world memory compaction algorithm is given, where multiple threads
compact the whole heap.  Compressor~\cite{Ker1} is a concurrent, parallel, and
incremental compaction algorithm which compacts the whole heap during a single
heap pass, achieving perfect compaction. A further parallel incremental
compaction approach is presented in~\cite{Ben1} where the heap is split into
pieces which are compacted one at a time by moving objects to a new memory
region. A fixup pass takes care of reference updates.  An algorithm with improved
compaction pause times via concurrent reference updates, using only half of the
heap, is given in~\cite{Oss1}. Each thread performs reference updates
proportional to its allocation requests. In~\cite{Kalibera09} the authors discuss
object replication versus forwarding pointer based compaction strategies. They
evaluate the performance in a non-concurrent virtual machine and show that
object replication may provide higher throughput there.

Garbage-collecting heap management systems that do provide response-time
guarantees on allocation and deallocation operations are Jamaica~\cite{Sie2}
as well as Metronome~\cite{Bac2}.  With Jamaica allocation and deallocation take
linear time in the size of the operation request.  Compaction is not needed
since memory objects do not occupy contiguous blocks of memory. Another garbage
collection approach based on non-contiguous memory allocation is discussed
in~\cite{Pizlo10} where memory access can be performed in constant time.
Metronome is a time-triggered garbage collector, which uses the same memory
layout as CF.  Compaction in Metronome is part of the garbage collection
cycles. The time used for compaction is estimated to at most 6\% of the
collection time~\cite{Bac1}, without precise guarantees. The performance of
Metronome depends highly on the mutator behavior.  MC$^2$~\cite{Sac2} is an
incremental soft real-time garbage collector designed for memory constrained
devices, which cannot provide hard guarantees on maximum pause time and CPU
utilization, but comes with low space overhead and tight space bounds.
Stopless~\cite{Piz1} is another garbage collector with soft guarantees on
response times. It provides low latency while preserving lock-freedom,
supporting atomic operations, controlling fragmentation by compaction, and
supporting multiprocessor platforms. The main contribution of Stopless is a
compaction algorithm which moves objects in the heap concurrently with program
execution. Exact bounds for response times, as well as fragmentation, are
missing in Stopless. Another incrementally compacting real-time garbage
collection algorithm where memory is divided into multiple pieces of equal
size, which get scavenged periodically resulting in bounded pause times is
presented in~\cite{Nilsen09}.  In~\cite{Kalibera11} the authors show
experimentally that the cost of handles in a real-time garbage collector is
negligible in comparison to implementations that do not use handles.
In~\cite{Bendersky11} the authors discuss worst-case fragmentation bounds for
different heap management strategies. Scheduling of garbage collection tasks
in real-time environments is discussed in~\cite{Schoeberl10}.

We remark that CF, like many of the above mentioned systems, is based
on segregated lists.  Approaches that are not based on segregated
lists, but rather on data structures which maintain locality of
objects, are known to perform better when accessing objects by
utilizing memory caches more effectively.  However, the use of
segregated lists enables providing and trading-off temporal and
spatial guarantees.

\section{Non-incremental Compact-fit}
\label{sec:non-inc-cf}

Compact-Fit (CF) is an explicit, dynamic heap management system that provides
strict temporal and spatial (fragmentation) guarantees.
Allocation as well as deallocation without compaction takes
constant time, whereas deallocation with compaction takes linear time
in the size of the object.


To be precise, there are two CF implementations~\cite{USENIX08}, but
in this article we only focus on the more fundamental so-called moving
implementation.

The set-up of CF is as follows: The memory is divided in pages of
equal size. Each page (in use) contains a certain number of
constant-sized page-blocks. In total there are finitely many available
page-block sizes, which determine to which size-class a page belongs
(namely all pages with a given page-block size belong to one
size-class). The pages are assigned to a size-class only if they are
used (non-empty). The number of page-blocks $\pi$ per page in a
size-class is therefore determined by the size of a page and the block
size. The state of a size-class depends on the state of the pages that
belong to it and is described by the values of the variable tuple
$$\langle h,n,u_1, \dots u_n\rangle$$ where $h$ is the total number of
allocated page-blocks in the size-class (its portion of the heap), $n$
is the number of not-full pages, and for each not-full page $i$, $u_i$
is the number of used page-blocks in the page.

An allocation request for an object of size $l$ is served by a page of
a best-fitting size-class. That is, for allocating an object a single
page-block is used in a page whose page-blocks are of the smallest
size still big enough to fit $l$. For example, if there are two
size-classes, one with page-blocks of size 10 and one with page-blocks
of size 20 units, then an allocation request for an object of size $l
\in \{11, 12, \dots 20\}$ will be served by a page of the size-class
20. If all pages in the best-fitting size-class are full, then a new
empty page is added to the size-class and the object is allocated in
this new page.

We allow for a constant number $\kappa>0$ of not-full pages per
size-class.  The aim in the design of CF is to control size-class
fragmentation, which is the space occupied by free page-blocks in
not-full pages (space not available for allocation in other
size-classes).  If deallocation happens, and the number of not-full
pages becomes $\kappa+1$ after this deallocation operation, then
compaction is invoked.  Compaction consists of moving a single object
from a not-full page to the page-block of the deallocated object,
which is the only empty page-block in that page.  As a result, after
compaction, the number of not-full pages in a size-class does not
exceed $\kappa$.

An object is assigned a unique abstract address (handle), which has to be
dereferenced whenever accessing an object field. This introduces a constant
object dereferencing overhead but facilitates predictability of reference
updates during compaction, i.e., whenever an object is moved in memory it
requires to update just its abstract address space entry.

\begin{figure*}
    \begin{center}
        \includegraphics[width=\textwidth]{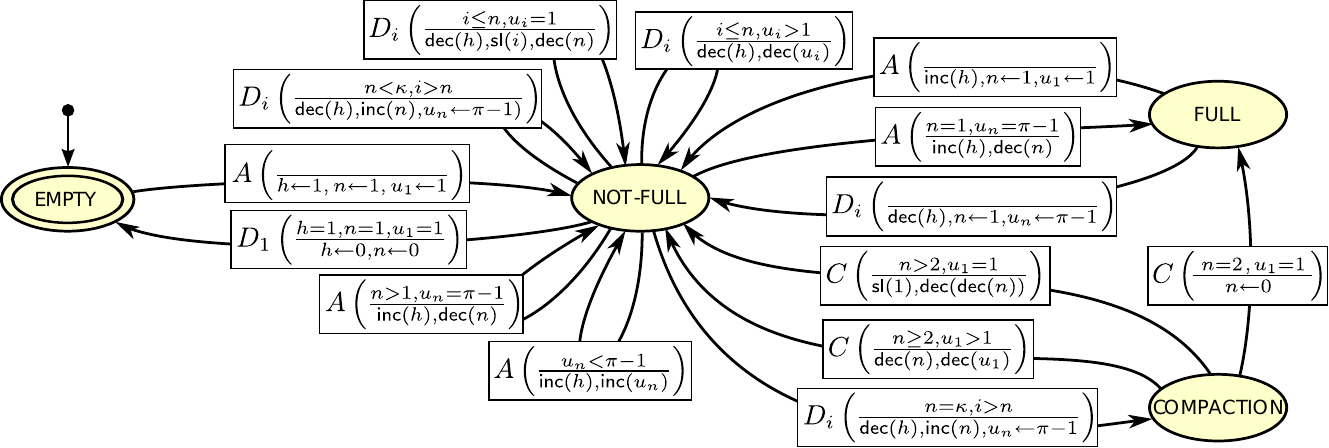}
        \caption{Size-class automaton with $\pi > 1$}
            \label{fig:sc-automaton}
    \end{center}
\end{figure*}

We show the CF algorithm in full detail in
Figure~\ref{fig:sc-automaton}, using a deterministic automaton, one
per size-class. For presentation purposes, we draw a quotient of the
state space of the size-class: {\scriptsize{\textsf{EMPTY}}} stands
for the single state $\langle 0,0\rangle$ representing an empty
size-class; {\scriptsize{\textsf{NOT-FULL}}} represents all states
with at least one not-full page where no compaction is needed, that is
$\langle h, n, u_1, \dots u_{n}\rangle$ with $0 < n \le \kappa$; the
state {\scriptsize{\textsf{FULL}}} represents all states with no
not-full pages and at least one full page, that is $\langle h,
0\rangle$ with $h > 0$; finally, {\scriptsize{\textsf{COMPACTION}}}
represents states $\langle h, \kappa+1, u_1, \dots, u_{\kappa+1}
\rangle$ in which compaction must be invoked.

The transitions in the automaton are labelled in the following way:
$A$ denotes allocation, $D_i$ deallocation in page $i$ (which may be
full or not-full, the latter is recognized by $i \le n$), and $C$
denotes a compaction step. Moreover, a transition fires if its premise
is satisfied, and results in a change of state described by its
conclusion. For updating a state, we use the operators $\leftarrow$
for assignment, $\textsf{dec}$ for decrement, $\textsf{inc}$ for
increment, and $\textsf{sl}$ for shift left. More precisely,
$\textsf{sl}(i)$ removes $u_i$ from a state
sequence, i.e., it changes a state $\langle h, n, u_1, \dots, u_{i-1},
u_{i}, u_{i+1}, \dots, u_{n}\rangle$ to the sequence $\langle h, n,
u_1, \dots, u_{i-1}, u_{i+1}, \dots, u_{n}\rangle$.

We explain several instructive transitions in full detail, and refer
the reader to Figure~\ref{fig:sc-automaton} for the full algorithm.\\



\noindent $A\left(\frac{}{h \leftarrow 1,\, n \leftarrow 1,\, u_1 \leftarrow
1}\right)$ from
{\scriptsize{\textsf{EMPTY}}} to {\scriptsize{\textsf{NOT-FULL}}}\\

\noindent This transition fires whenever allocation is
requested in the
empty state. As a result the state changes to $\langle 1, 1,1\rangle$.\\


\noindent $D_i\left(\frac{ i\,\le\, n,\, u_i\, > \,1}{\textsf{dec}(h),\,\textsf{dec}(u_i)}\right)$
from
{\scriptsize{\textsf{NOT-FULL}}} to {\scriptsize{\textsf{NOT-FULL}}}\\

\noindent This transition is taken upon a deallocation step
in a not-full page which remains non-empty after the
deallocation. The change in the state is that the number of used page-blocks is decremented by 1, and, as in every deallocation step, the heap
size decreases by 1. \\


\noindent $A\left(\frac{}{\textsf{inc}(h),\,n \leftarrow 1,\, u_1 \leftarrow
1}\right)$ from
{\scriptsize{\textsf{FULL}}} to {\scriptsize{\textsf{NOT-FULL}}}\\

\noindent Whenever an object is allocated in a state of
the class {\scriptsize{\textsf{FULL}}} a new empty page has to be
added to the size-class, and allocation happens in this page. As a
result this new page becomes the only not-full page of the size-class
with a single page-block used. The value of $h$ increases by one, as
with any allocation operation.\\


\noindent $D_i\left(\frac{n = \kappa,\, i > n}{\textsf{dec}(h),\textsf{inc}(n), u_n
\leftarrow \pi-1}\right)$ from
{\scriptsize{\textsf{NOT-FULL}}} to {\scriptsize{\textsf{COMPACTION}}}\\

\noindent With this transition we are in a situation when
after the required deallocation operation, in the $i$-th page which was
full, we have more than $\kappa$ not-full pages. Therefore, compaction must
be invoked in the next step.\\


\noindent $C\left(\frac{u_1 = 1, \, n > 2}{\textsf{sl}(1), \,
\textsf{dec}(\textsf{dec}(n))}\right)$ from
{\scriptsize{\textsf{COMPACTION}}} to {\scriptsize{\textsf{NOT-FULL}}}\\

\noindent Being in state
{\scriptsize{\textsf{COMPACTION}}}, the next transition has to be of
type $C$. Moreover, note that $n = \kappa+1 \ge 2$. During the compaction
step a page-block is moved from the first not-full page (represented by
$u_1$) to the last not-full page, namely the one in which deallocation
just happened. This particular transition fires if the first not-full
page has just one page-block. As a result it becomes empty after the
transition, whereas the last not-full page becomes full. Since $n > 2$
the transition leads to the state ${\scriptsize{\textsf{NOT-FULL}}}$.
The operation shift left is needed to remove the value $u_1$ for the now
empty page.

\begin{figure}
    \begin{center}
        \includegraphics[width=.75\textwidth]{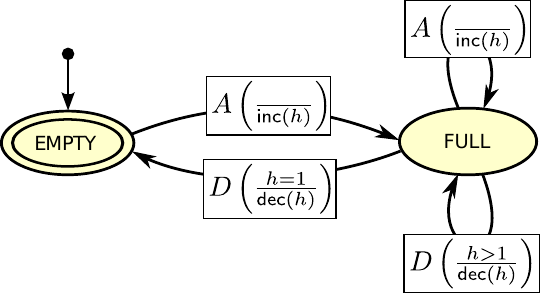}
        \caption{Size-class automaton with $\pi = 1$}
            \label{fig:sc-simple-automaton}
    \end{center}
\end{figure}
We note that in case $\pi = 1$, i.e., in a size-class in which each page
consists of exactly one page-block, there are no not-full pages. A page is
either empty or full. In this case compaction can never happen.
Therefore, the size-class automaton simplifies significantly as shown in
Figure~\ref{fig:sc-simple-automaton}.

We have chosen the automaton presentation of CF in order to prepare
the ground for the concurrent version. For the original presentation
of CF, we refer the interested reader to~\cite{USENIX08}.  We extend
the non-incremental CF with blocking and non-blocking synchronization
mechanisms so that multiple threads can share a single (or multiple)
instance(s). In particular, we make the size-class automaton
transitions (including a combination of a deallocating transition
followed by a compacting step) atomic. As a result, multiple threads
can execute and use CF in parallel, interleaving between the atomic
transitions. The details of the particular implementation and the
various choices of synchronization mechanisms are discussed in
Section~\ref{sec:impl}. The results are encouraging for throughput
oriented environments, see Section~\ref{sec:exp}.

%
%
%
%
%

\section{Probabilistic Analysis}
\label{sec:prob}

We present a probabilistic analysis of CF which shows that, for
{particular} mutator behavior, both compaction and worst-case
size-class fragmentation are less likely to happen with increasing
partial compaction bounds~$\kappa$.  Compaction may therefore be set
to large $\kappa$ or even turned off if guarantees on memory
fragmentation are not required.

{We conjecture that compaction in CF may actually be turned off
  in some applications while maintaining bounded
  memory fragmentation with high probability.  The probabilistic
  analysis of CF we present here is not a complete proof of this
  conjecture but nevertheless motivates non-compacting CF and points
  in the direction of potential solutions outside the scope of this
  paper which will need to involve representative classes of mutator
  behavior.}
Interestingly, it is the partitioned memory layout of CF that allows
for such an analysis, since the partitioning into pages and
size-classes significantly reduces the state space of the model.
Other memory allocators may not allow such an analysis.

We aim at answering the following two questions:
\begin{list}{\labelitemi}{\leftmargin=1.2em}
\item[1.] What is the probability that compaction happens?
\item[2.] What is the probability of  worst-case fragmentation?
\end{list}

We analyze the behavior of CF given a mutator, which is
a sequence of allocations $A$ and deallocations $D$, hence a word in $\{A,D\}^*$. A mutator is
not aware of the internal CF configuration, e.g. in which page deallocation happens.
Therefore, we abstract away from the index $i$ in the deallocation label $D_i$ and the CF size-class
automaton becomes a probabilistic I/O automaton (PIOA)\footnote{{The full definition of a PIOA is out of scope of this paper, instead of giving the general definition we describe the concrete CF size-class automaton as a PIOA.}}~\cite{WSS97:tcs}, with input actions $A$ and $D$
provided by the mutator, and an output action $C$ provided by CF. The states of this automaton
are either input states
in which $A$ and $D$ are enabled, or output states in which $C$ is enforced, which makes
it simpler than general PIOA. {In an input state, each input action leads to a discrete probability distribution over possible next states. Hence, in an input state, if $A$ happens, we reach a next state with a given probability, and the sum of the probabilities after $A$ equals 1. Symmetrically, after $D$ we reach a next state with a given probability and the sum of the probabilities after $D$ equals 1. In an output state, the single output action $C$ happens with probability 1. }
For brevity we only discuss in detail the behavior of a
single state. In a state $\langle h,n,u_1, \dots, u_n\rangle$ with $n \le \kappa$, upon deallocation $D$,
there are several possible next states that are reached with different probabilities: for all $i$ with
$u_i > 1$, with probability $\frac{u_i}{h}$ deallocation happens in the not-full page
$i$ which will remain not-full afterwards and the next state becomes $\langle h-1, n, u_1, \dots, u_{i-1},
u_i - 1, u_{i+1}, \dots, u_n\rangle$; for all $i$ such that $u_i = 1$ with probability $\frac{1}{h}$
deallocation happens in page $i$ reducing the number of not-full pages and the
next state is $\langle h-1, n-1, u_1, \dots, u_{i-1}, u_{i+1}\dots, u_n\rangle$; and
with probability $\frac{h - \sum_i u_i}{h}$ the next state is $\langle h-1, n+1, u_1, \dots, u_n, \pi-1 \rangle$
as deallocation happens in a full page.
The allocation and compaction transitions remain the same as in the deterministic automaton, they
happen with probability 1 in states in which they are enabled.
This way we get the \emph{full} PIOA model, with initial state $\langle 0,0\rangle$.

The full PIOA model together with a mutator induces a discrete-time Markov chain, the \emph{full} DTMC,
by pruning out
the allocation/deallocation possibilities that the mutator does not prescribe in each state and
abstracting away from the transition labels. The full DTMC model results in a large
state space already for small values of $h$, $\pi$, and $\kappa$.

To reduce the number of states, we
consider only mutators of the shape
$A^hD^d$ which perform $h$ allocations followed by $d$ deallocations.
We analyze portions of the full model by setting the state reached
after performing $h$ allocations as initial state. This is the state
 $\langle h,0\rangle$ if $h \mod \pi = 0$, or
$\langle h, 1, h\mod \pi\rangle$ otherwise. Then we consider the portion of
the full model reachable in $d$ deallocations. We refer to $d$ as the \emph{deallocation level}.
Even such versions of the full model are too big:
for $h=80$, $\pi=10$, and $\kappa=5$ the DTMC model in Prism~\cite{Prism}
has 1429506 states and 2818395 transitions, and for $h=80$, $\pi=10$, and $\kappa=6$,
Prism runs out of memory.


\begin{figure*}
\begin{center}
\subfigure[compaction\label{fig:prob-comp}]{
\includegraphics[width=\textwidth]{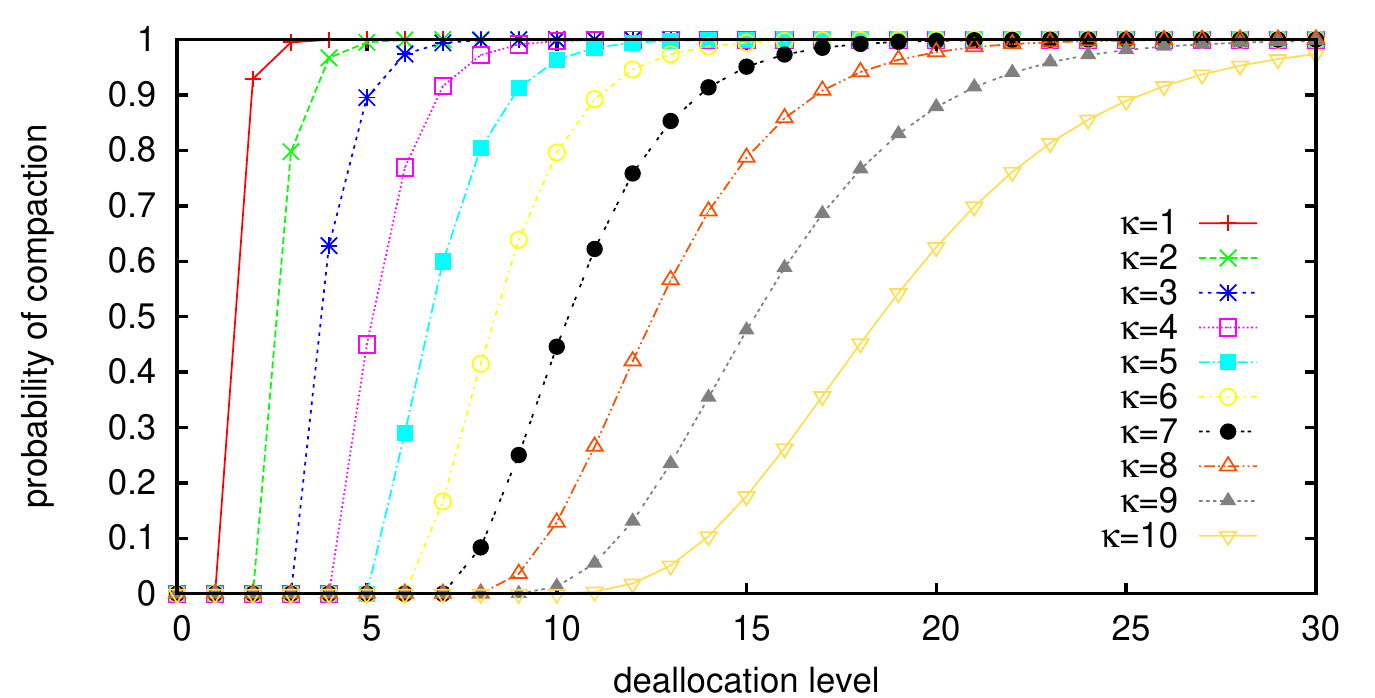} }
\subfigure[worst-case fragmentation\label{fig:prob-frag}]{
\includegraphics[width=\textwidth]{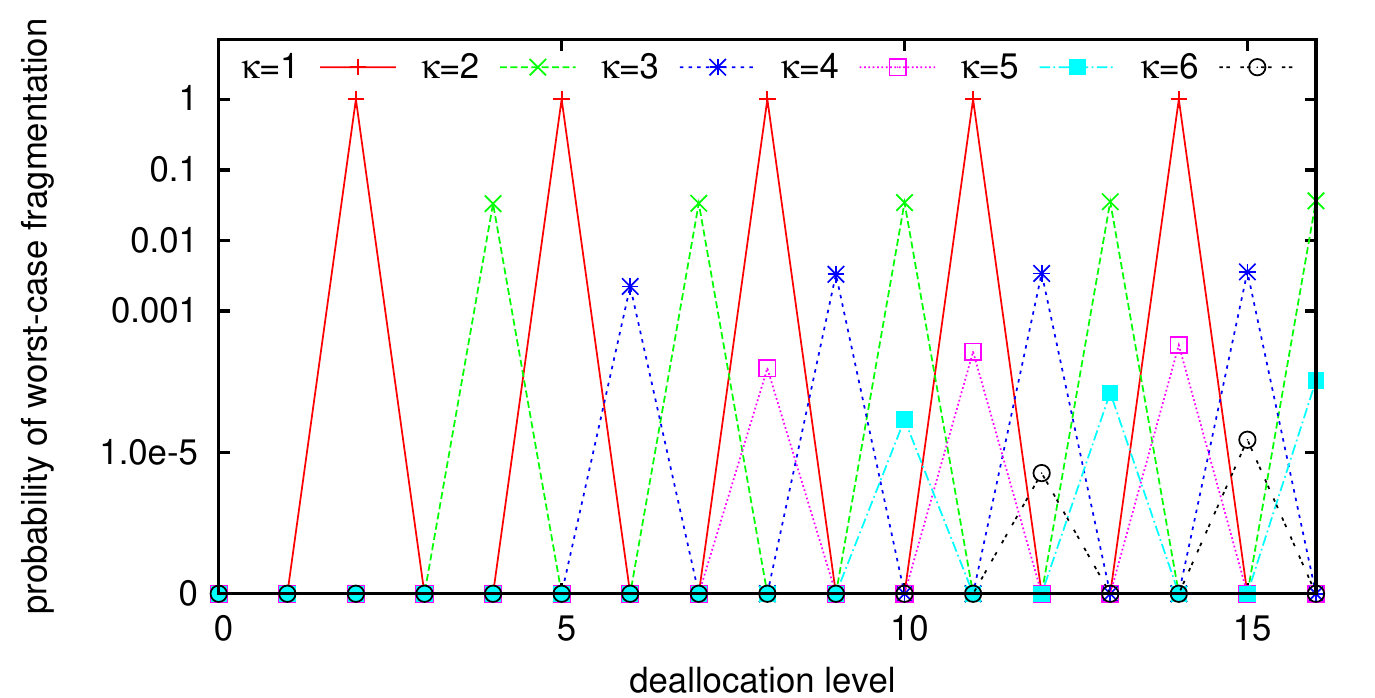}}
\caption{Probability of reaching compaction and worst-case fragmentation}
\end{center}
\end{figure*}

The probability of compaction
is the probability of reaching a compacting state, i.e.,
a state with $n = \kappa +1$. The probability of reaching a specified state
in a DTMC is  the sum over all paths of the probability of
reaching the state along a path, where the probability of reaching the
state along a path is calculated as the product of the probabilities on the
path until the state is reached, or equals zero if the state is not reached (in at most $d$ steps).
We run both Prism and our own program for exact calculation of the probability of compaction
 on the model
(in Prism only as long as no state space explosion occurs).\footnote{{Prism is a general multi-purpose probabilistic model checker applicable to many different models. As all model checkers it suffers from state space explosion. We have implemented a simple single-purpose program that calculates the probabilities of the full DTMC as described above. The results coincide with Prism, but with our simple program we were able to calculate the probabilities on our simple models for larger models than with Prism.}}
The results of
our exact calculations and of Prism coincide, and they are presented
in Figure~\ref{fig:prob-comp} for the values $h = 1400$, $\pi = 100$,
and varying values of $\kappa$ and $d$. The particular values of $h$ and $\pi$ are not significant,
we have chosen them so that the probability graphs are sufficiently apart from each other.
As expected, the probability of reaching a state where
compaction happens for a fixed $\kappa$ increases with increasing $d$, and
it overall decreases when increasing $\kappa$.

Given a state $\langle h, n, u_1, \dots, u_n\rangle$ the (size-class) fragmentation in this state is calculated as
$F = n\cdot\pi - \sum_{i=1}^n u_i.$ The probability  of worst-case fragmentation is the probability of reaching a worst-case fragmentation
state, i.e., a state with fragmentation $F = \kappa\cdot(\pi - 1)$.
The results are shown in Figure~\ref{fig:prob-frag} for $h = 120$, $\pi = 3$,
 and varying values of $\kappa$ and $d$. We present the results for small values of $\pi$ so that the effect of emptying a
page within $d$ deallocations can be seen even for small values of $d$.
 Given a partial compaction bound $\kappa$, the probability of
worst-case fragmentation oscillates periodically as $d$ increases, reaching a maximum value for certain values of $d$.
This maximal probability of worst-case fragmentation decreases with increasing $\kappa$, as intuitively expected.
Note that the $y$-axes has a logarithmic scale, and the maximum probabilities of worst-case fragmentation are
very low, for $\kappa > 1$.

%
%
%

\section{Incremental Compact-fit}
\label{sec:inc-cf}

For applications which require low latency and run on
memory-constrained systems, we provide an extension of CF that allows
for incremental compaction, i.e., incremental moving of a single
object.

The incremental extension of CF performs compaction, i.e., moving of a
single object, by an incremental moving operation. The reason why
compaction is made incremental is its dominating linear complexity.
This incremental extension is the first step towards a design of
latency-efficient concurrent CF.  For a concurrent incremental version
of CF, allocation, deallocation, and incremental compaction are made
atomic, leaving space for other interleaving threads between the
atomic steps. As a result the waiting times of concurrent threads, and
therefore their response times, decrease, although the compaction
throughput may also decrease.

\begin{figure}[ht!]
    \begin{center}
        \includegraphics[width=0.5\textwidth]{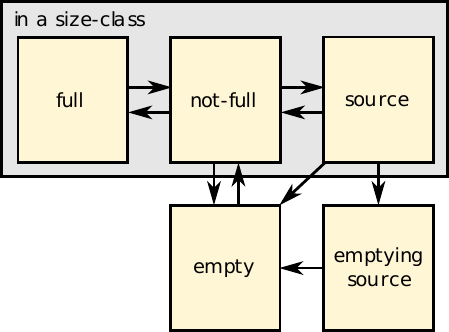}
        \caption{The lifetime of a page}
        \label{fig:page-lifetime}
    \end{center}
\end{figure}

There is a global fixed compaction increment $\iota>0$ which
determines the portion of a page-block being moved in an incremental
step. The value of $\iota$ may even be larger than some page-block
sizes, in which case the whole compaction operation is done
non-incrementally, in one step. We refer to a page-block under
incremental moving as the \emph{source page-block}, and the page-block
to which the object is moved as the \emph{target page-block}.  The
state of each size-class and its administration gain complexity in the
incremental version.  In a size-class, apart from the full and
not-full pages, there may exist one \emph{source page}. In a source
page there are used page-blocks and source page-blocks. The latter are
page-blocks that are in the process of being incrementally moved. One
source page suffices, since compaction in CF requires moving a used
page-block which is now always taken from the source page. Allocation
never happens in a source page. A source page always contains at least
one used page-block.  If a source page looses all its used page-blocks
(due to deallocation or compaction), it is removed from the size-class
and placed into a global pool $E$ of \emph{emptying source pages}.
All pages in the pool contain page-blocks that are involved in ongoing
incremental compaction operations.  The space occupied by source
page-blocks and free page-blocks in (emptying) source pages, which is
(temporarily) not available for allocation in any size-class, is
called transient size-class fragmentation.  When all incremental
compaction operations in an emptying source page finish, then the page
is returned to the global list of free pages.  On the other hand, if
all incremental compaction operations within a source page finish,
i.e., the source page has no more source page-blocks, and if there are
still used page-blocks in the source page, then there are two
possibilities: (1) the source page becomes a not-full page, if the
number of not-full pages is smaller than the partial compaction bound,
or (2) the source page is kept as a potential source page without
source page-blocks, otherwise. The evolution of a page is shown in
Figure~\ref{fig:page-lifetime}.


The state of a size-class is described by a tuple
$$\langle h, n, u_1, \dots, u_{n}, u_s, s, m_1, \dots, m_s\rangle$$
where, as before, $h$ denotes the current heap size, $n$ is the number
of not-full pages such that $n \le \kappa + 1$ with $\kappa$ being the
partial compaction bound, and the values of $u_1, \dots, u_{n}$ are
the numbers of used page-blocks in the not-full pages, respectively.
The value of $u_s$ equals the number of used page-blocks in the source
page, with $u_s = 0$ representing that there is no source page in the
size-class.  The variable $s$ contains the number of source
page-blocks in the source page and equals 0 if there is no source
page. Note that $s = 0$ and $u_s >0$ represents the existence of a
potential source page, as discussed above. Finally, $m_1, \dots, m_s$
are the sizes of the portions of the $s$ source page-blocks that have
already been moved.

\begin{figure*}
    \begin{center}
        \includegraphics[width=1.1\textwidth]{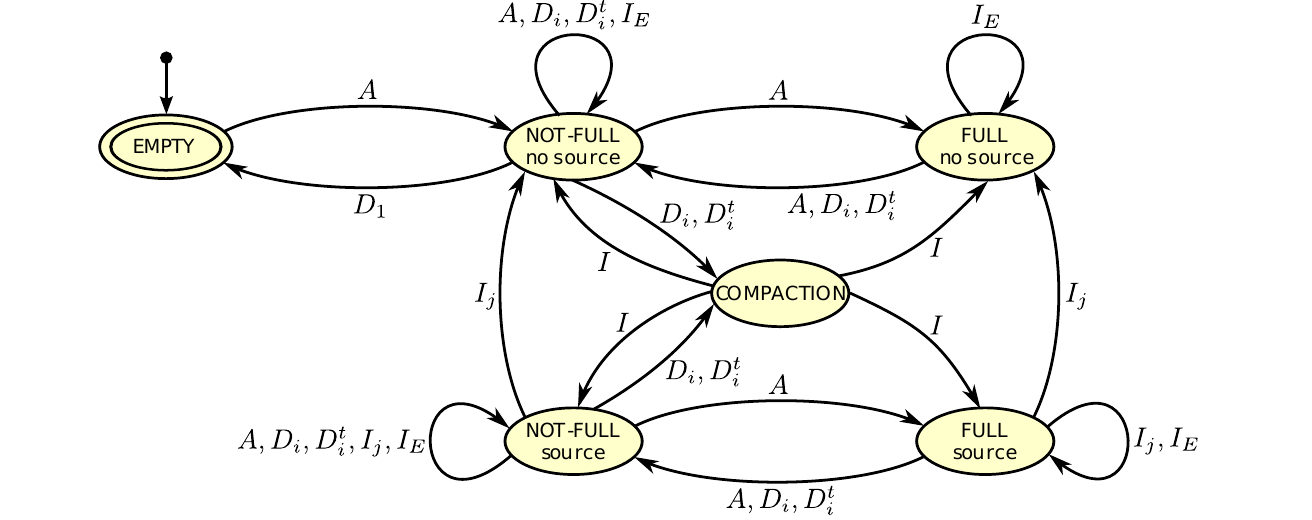}
        \caption{Incremental size-class automaton with $\pi > 1$}
            \label{fig:sc-inc-automaton}
    \end{center}
\end{figure*}

Figure~\ref{fig:sc-inc-automaton} shows an abstraction of the
size-class behavior.  Similar to Figure~\ref{fig:sc-automaton}, we use
abstract states to describe the state changes:
{\scriptsize{\textsf{EMPTY}}} stands for the single state $\langle
0,0,0,0\rangle$ representing an empty size-class; the state
{\scriptsize{\textsf{NOT-FULL, no source}}} represents all states with
at least one not-full page where no compaction is needed and no source
page is present, that is $\langle h, n, u_1, \dots u_{n}, 0, 0\rangle$
with $0 < n \le \kappa$; the state {\scriptsize{\textsf{FULL, no
      source}}} represents all states with no not-full pages, at least
one full page, and no source page, that is $\langle h, 0, 0, 0\rangle$
with $h > 0$; {\scriptsize{\textsf{NOT-FULL, source}}} represents all
states with at least one not-full page where no compaction is needed
and a source page, that is $\langle h, n, u_1, \dots u_{n}, u_s, s,
m_1, \dots, m_s\rangle$ with $0 < n \le \kappa$, $u_s > 0$;
{\scriptsize{\textsf{FULL, source}}} represents all states with no
not-full pages, at least one full page, and a source page, that is
$\langle h, 0, u_s, s, m_1, \dots, m_s\rangle$ with $h > 0$ and $u_s >
0$; finally, {\scriptsize{\textsf{COMPACTION}}} is used to represent
states $\langle h, \kappa+1, u_1, \dots, u_{\kappa+1}, u_s, s, m_1,
\dots, m_s \rangle$ in which compaction must be invoked.  We note that
the automaton and the discussion in this section is under the
assumption that the number of page-blocks in a page is larger than 1,
$\pi > 1$. The degenerate case with $\pi = 1$ is of no interest.

A state change in a size-class happens upon allocation ($A$),
deallocation ($D_i, D_i^t$), or incremental compaction ($I$, $I_j$,
$I_E$) transitions.  
A transition $I$ represents an initial
incremental compaction step, $I_j$ is any further incremental
compaction step which involves a source page, and $I_E$ is a further
incremental compaction step which involves an emptying source page.

We next present the actual changes of states in a size-class in full
detail upon allocation, deallocation, and incremental compaction.\\

 \noindent{\bf{Allocation.}}
Allocation steps are the same as in the non-incremental automaton
since the source page is not influenced by allocation.  In detail, in
a state $\langle h, n, u_1, \dots, u_{n}, u_s, s, m_1, \dots,
m_s\rangle$ there are three cases:
\begin{list}{\labelitemi}{\leftmargin=1.2em}
\item[1.] If $n = 0$, that is, there are no not-full pages, then after allocation $h$ increases by 1,
$n$ becomes 1, and $u_1$ becomes 1.
\item[2.] If $0 < n \le \kappa$ and $u_{n} < \pi-1$, that is, there is a not-full page and after an allocation it will not get full,
then both $h$  and $u_{n}$ increase by 1.
\item[3.] If $0 < n \le \kappa$ and $u_{n} = \pi-1$, that is, a not-full page will get full, then $h$ increases by 1 and $n$ decreases
by 1. Note that this may change a state from ``not-full'' to ``full'' in case $n=1$.
\end{list}
Allocation is not possible in a ``compaction'' state, i.e., a state with $n = \kappa+1$.\\

\noindent{\bf{Deallocation.}}
We distinguish two types of deallocation steps denoted by $D_i$ and
$D_i^t$. A step $D_i$ denotes deallocation in page $i$ where the
deallocated page-block is not a target of an ongoing incremental
moving. In contrast, $D_i^t$ denotes deallocation in page $i$ of a
page-block which happens to be a target of an ongoing incremental
moving. If $i = 0$, then deallocation happens in the source page;
if $1 \le i \le n$, then deallocation happens in one of the not-full
pages; and if $i > n$ a page-block is deallocated in a full page.

Similar to the non-incremental CF, the change of state after $D_i$ can
be described by the following cases:
\begin{list}{\labelitemi}{\leftmargin=1.2em}
\item[1.] If $1 \le i \le n \le \kappa$ and $u_{i} > 1$, or if $i = 0$, $u_s > 1$, and $n \le \kappa$,
i.e., deallocation happens in a not-full or source page which will not get empty(ing), then $h$ decreases
by 1 and either $u_{i}$ or $u_s$ decreases by 1, respectively.
\item[2.] If $1 \le i \le n \le \kappa$ and $u_{i} = 1$, i.e., deallocation happens
in a not-full page which becomes empty afterwards, then both $h$ and $n$ decrease by 1, and the variable $u_{i}$
is removed from the
state.
\item[3.] If $i = 0$, $u_s = 1$, and $n \le \kappa$, i.e., deallocation happens in a
source page which becomes emptying afterwards, then the source page is moved to the pool of emptying source pages $E$ and
both $s$ and $u_s$ are set to 0. As a result the size-class
does not have a source page.
\item[4.] If $i > n \le \kappa$, which means that deallocation happens in a full page, then $h$ decreases by 1, $n$ increases by 1,
 and $u_{n}$
gets the value $\pi-1$. If originally $n = \kappa$, then this step triggers a compaction operation.
\end{list}

In addition, there are four cases describing the change of state after
$D_i^t$ steps.  They correspond to the cases for $D_i$ except that at
the end of such a step the ongoing incremental compaction operation to
the deallocated target page-block is canceled, the target page-block is deallocated, and the source
page-block is deallocated. Hence, the (canceled) ongoing compaction
operation finishes earlier than it normally would. We refer to the
situation when a thread performs a $D_i^t$ step as a
\emph{deallocation conflict}.

Deallocation is also not possible in a ``compaction'' state with $n = \kappa+1$.\\

\noindent{\bf{Incremental compaction.}}
Incremental compaction is triggered in case $n = \kappa+1$, just like
compaction is triggered in the non-incremental CF. In addition, there
may be incremental compaction steps involving emptying source pages
from any other state, and incremental compaction steps involving the
source page from any state with a source page.

In a state $\langle h, \kappa + 1, u_1 \dots, u_{{\kappa + 1}}, u_s, s, m_1,
\dots, m_s\rangle$ an initial incremental compaction step is the only
possible step. Note that in such a state $u_{{\kappa + 1}} = \pi-1$ since the
previous step was a deallocation in a full page.  We refer to this
unique free page-block in the last not-full page as \textsf{tb}.  The
initial incremental compaction step must be atomic together with the
preceding deallocation step. We use $\beta$ to denote the
size of page-blocks in the size-class. We have the following cases:
\begin{list}{\labelitemi}{\leftmargin=1.2em}
\item[1.] If $u_s = 0$, meaning that there is no source page in the size-class, then since $n = \kappa+1 \ge 2$
the first page becomes the new (potential) source page, i.e., $u_s$ is assigned the value of $u_1$, $s$ becomes 0, $n$ decreases by 1,
and $u_1$ is removed
from the state. After this, the state is no longer a ``compaction'' state.
\item[2.] If $u_s > 0$, then a source page-block \textsf{pb} is to be moved to \textsf{tb}. There are two possible cases:
\begin{list}{\labelitemi}{\leftmargin=1.2em}
\item[-] The page-block \textsf{pb} is not a target page-block of an ongoing incremental moving operation. In this case
there are two subcases representing an initial incremental compaction step: (1) if $\iota < \beta$, in which
case the compaction operation needs more than just one step, then
$u_s$ decreases by 1, $s$ increases by one, $m_s$ is assigned the value of $\iota$ and a portion of size $\iota$ is moved from \textsf{pb} to
\textsf{tb}; (2) if $\iota \ge \beta$, then the whole \textsf{pb} is moved to
\textsf{tb} in one step and $u_s$ decreases by 1.
\item[-] The page-block \textsf{pb} is a target of a (unique) ongoing incremental operation from a source
page-block \textsf{sb}. In this case we are in a situation
of a \emph{compaction conflict}. Note that \textsf{sb} must be in an emptying source page in $E$. Then the ongoing
incremental moving operation from \textsf{sb} to \textsf{pb} is canceled, \textsf{pb} is deallocated,
and a new initial incremental moving
operation starts from \textsf{sb} to \textsf{tb}. Again $u_s$ decreases by 1.
\end{list}
In any case, $n$ decreases by 1. In case $u_s = 0$, the source page becomes emptying, it is moved to
the pool of emptying source pages $E$, and $s$ becomes 0.
\end{list}
Note that the chosen way to resolve the compaction conflict is crucial for bounded compaction response times, since
it avoids transitive compaction chains. Namely, a compaction conflict ends an existing compaction and starts a new one, so
the duration of a particular compaction operation may only decrease due to a compaction conflict.

In addition, there are three more cases for a change of state due to an ongoing incremental compaction step $I_j$,
where $j$ is an index of a source page-block in the source page that the incremental compaction
step applies to. In a state $\langle h, n, u_1 \dots, u_{n}, u_s, s, m_1, \dots, m_s\rangle$ where $I_j$
is applicable, i.e., $u_s > 0$ and $s \ge j$, after an incremental compaction step $I_j$ we have:

\begin{list}{\labelitemi}{\leftmargin=1.2em}
\item[3.] If $m_j + \iota < \beta$, then $m_j$ is incremented by $\iota$, i.e., another portion of the source page-block gets
copied to the target page-block.
\item[4.] If $m_j + \iota \ge \beta$ and $s > 1$, i.e., this is the last incremental step for the compaction operation
which still keeps the source page, then the
number of source page-blocks $s$ decreases by 1, the variable $m_j$ is removed from the state.
\item[5.] If $m_j + \iota \ge \beta$ and $s = 1$, i.e., the compaction operation finishes after this incremental step
and the source page will no longer exist in the size-class, then
$s$ gets the value 0. Furthermore, the source
page either becomes a not-full page if $n < \kappa$ (in which case $n$ increases by 1, $u_n$ is assigned the value of $u_s$,
$u_s$ becomes 0) or it is kept as a potential source page.
\end{list}

Finally, there is a possibility for incremental operations $I_E$ which
do not change the state, but only change the global pool $E$ of
emptying source pages. We skip the details on the description and the
update of $E$ due to $I_E$ operations.


We remark that the behavior of any thread can be expressed by a
sequence of allocations and deallocations.  If a deallocation triggers
compaction, then before the thread can continue with any other
allocation or deallocation operation all incremental steps needed for
the compaction must be finished. The first of these steps is an
initial incremental compaction step $I$ which may be an initial
incremental moving step in case of compaction conflict. If it is the
case, then all other incremental steps are of type $I_E$. Otherwise,
if there is no compaction conflict, a sequence of $I_j$ incremental
steps will be performed, and in case the source page becomes emptying
a sequence of $I_E$ incremental steps, in order to complete the
compaction operation.

\section{Complexity vs. Fragmentation}
\label{sec:complexity}

\newlength{\longline}
\settowidth{\longline}{xxxxxxxxxx}

\begin{table*}
\small
\centering
\begin{tabular}{|l|c|c|c|c|c|}
\hline
& malloc & free & latency & memory size & \parbox[t]{\longline}{size-class \\ fragmentation} \\
\hline
$1$-CF$(\infty,\infty)$ & $O(n)$ & $O(n)$ & $O(1)$ & $O(n*m*\pi*\beta)$ & \parbox[t]{\longline}{$O(n*m*$ \\ \text{~~~}$(\pi-1)*\beta)$} \\
\hline
$1$-CF$(\kappa,\infty)$ & $O(n)$ & $O(n+\beta)$ & $O(\beta)$ & \parbox[t]{\longline}{$O((n*m+$ \\ \text{~~~}$\kappa*(\pi-1))*$ \\ \text{~~~}$\beta)$} & \parbox[t]{\longline}{$O(\kappa*(\pi-1)*$ \\ \text{~~~}$\beta)$} \\
\hline\hline
$n$-CF$(\infty,\infty)$ & $O(1)$ & $O(1)$ & $O(1)$ & $O(n*m*\pi*\beta)$ & \parbox[t]{\longline}{$O(n*m*$ \\ \text{~~~}$(\pi-1)*\beta)$} \\
\hline
$n$-CF$(\kappa,\infty)$ & $O(1)$ & $O(\beta)$ & $O(\beta)$ & \parbox[t]{\longline}{$O(n*(m+$ \\ \text{~~~}$\kappa*(\pi-1))*$ \\ \text{~~~}$\beta)$} & \parbox[t]{\longline}{$O(n*\kappa*$ \\ \text{~~~}$(\pi-1)*\beta)$} \\
\hline\hline
$1$-CF$(\kappa,\iota)$  & $O(n)$ & \parbox[t]{\longline}{$O(n+\beta+$ \\ \text{~~~}$\lfloor\frac{\beta}{\iota}\rfloor)$} & $O(\min(\beta,\iota))$ & \parbox[t]{\longline}{$O((n*m+$ \\ \text{~~~}$n*\pi+$ \\ \text{~~~}$\kappa*(\pi-1))*$ \\ \text{~~~}$\beta)$} & \parbox[t]{\longline}{$O((n*\pi+$ \\ \text{~~~}$\kappa*(\pi-1))*$ \\ \text{~~~}$\beta)$} \\
\hline
\end{tabular}
\caption{Time complexity of malloc and free as well as worst-case
  system latency, memory size, and size-class fragmentation per CF
  configuration and size-class}
\label{tab:complexity}
\end{table*}

Table~\ref{tab:complexity} shows the time complexity of malloc and
free as well as the worst-case system latency, memory size, and
size-class fragmentation per CF configuration with $n$ threads and $m$
per-thread-allocated page-blocks in a size-class with $\pi$
page-blocks of size $\beta$ per page.  The fragmentation caused by
partitioning memory~\cite{Bac1,USENIX08} is not considered here.
Although the partial compaction bound $\kappa$ and the compaction
increment $\iota$ are kept constant in our current implementations,
both $\kappa$ and $\iota$ may be changed dynamically at runtime, which
is an interesting topic for future work.  System latency is here the
portion of the delay a thread may experience, from invoking malloc or
free until the operation actually begins executing, caused by
currently executing, non-preemptive CF operations, not including the
synchronization overhead. {Recall that $i$-CF$(\kappa,\iota)$ denotes a CF configuration
where $i$ instances of concurrent CF run in parallel with partial compaction bound $\kappa$ and compaction increment~$\iota$.
If $\iota = \infty$, then incremental compaction is turned off. If $\kappa = \infty$, then compaction (and hence also incremental compaction) is turned off.}

Since all operations of the non-compacting $1$-CF$(\infty,\infty)$
configuration take constant time, the complexity of malloc and free
only depends linearly on the number of competing threads assuming fair
scheduling.  System latency is bounded by a constant.  However, the
worst case in memory consumption is one page for each allocated object
due to potentially high size-class fragmentation, which has
asymptotically the same bound as the overall memory consumption.  The
compacting $1$-CF$(\kappa,\infty)$ configuration trades-off complexity
of free and worst-case latency for better bounds on memory consumption
by limiting size-class fragmentation through partial compaction.  Note
that in this case size-class fragmentation is independent from the
number of threads and allocated objects.

The results for the $n$-CF configurations, in particular the worst
cases in memory size and size-class fragmentation, as shown here, are
obtained under the assumption that there is no sharing among the $n$
CF instances.  The time complexity of malloc and free of both
multiple-instance configurations goes up to the respective
single-instance cases if there is sharing among the $n$ CF instances.
While the non-compacting $n$-CF$(\infty,\infty)$ configuration
requires in the worst case no more memory than the non-compacting
single-instance configuration, the compacting $n$-CF$(\kappa,\infty)$
configuration actually does require in the worst case more memory than
the compacting single-instance configuration since partial compaction
is performed per instance.  However, allocation and deallocation
throughput may increase with both multiple-instance configurations
with a decreasing degree of sharing among the $n$ CF instances
(without an increase in worst-case system latency).

The incremental $1$-CF$(\kappa,\iota)$ configuration actually improves
the worst case in system latency whenever the compaction increment
$\iota$ is less than the page-block size of the size-class with the
largest page-blocks, at the expense of the complexity of free through
more preemptions and at the expense of memory consumption through
increased transient size-class fragmentation.  In comparison to the
non-incremental, compacting $1$-CF$(\kappa,\infty)$ configuration,
there may be up to $n$ additional (emptying) source pages in the
system where $n$ is the number of threads.  The worst case in
non-transient size-class fragmentation does not increase.

\section{Implementation}
\label{sec:impl}

Sequential CF~\cite{USENIX08} uses three data structures to manage its
heap: {abstract address}, {page}, and {size-class}.  Additionally,
empty pages and available abstract addresses are organized in global
LIFO lists.

An abstract address is a forwarding pointer word.

A {page} contains a page header holding the meta data of the page and
the storage space into which objects are allocated.  The size of each
{page} is~16KB.  All pages are kept aligned in memory.  The page
header consists of: two pointers used to insert the page into a
doubly-linked list, a counter of allocated page-blocks in the page, a
reference to the size-class of the page, and a bitmap where each set
bit represents a used page-block in the storage space.  The bitmap is
used for fast location of free and used blocks.

A size-class contains two doubly-linked lists of pages which store the
full and the not-full pages, respectively, and a counter of the number
of not-full pages.

Global data structures are used to organize data structures which do
not belong to a particular size-class.  Such are a LIFO list of empty
pages and a LIFO list of free abstract addresses.  The implementation
details that make these data structures concurrent and scalable will
be discussed in the following subsections.

\begin{figure*}
    \begin{center}
        \includegraphics[width=1.0\textwidth]{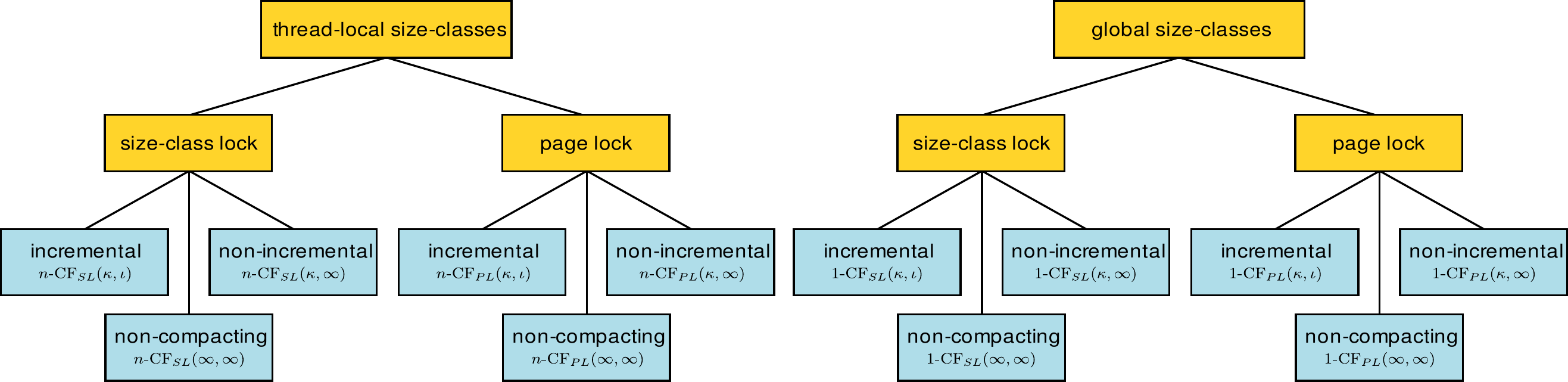}
        \caption{Concurrent CF versions}
        \label{fig:cf-versions}
    \end{center}
\end{figure*}

Figure~\ref{fig:cf-versions} presents an overview of all implemented
CF versions (leafs of the tree) and introduces terminology.

\subsection{Concurrent Non-incremental CF}\label{subsec:conc-ninc-cf}

We use blocking and non-blocking mechanisms to allow for concurrent
use of CF by multiple threads.  In particular, locks are used to make
the size-class automaton transitions atomic (allocation, deallocation
that does not cause compaction, and deallocation with compaction) and
non-blocking mechanisms are used to render access to the global LIFO
lists atomic and scalable.

In all concurrent implementations size-classes are kept 128B aligned
in memory, to avoid cache conflicts of concurrent threads.

We implement locks at two possible levels: size-class locks and page
locks.  The choice of lock level is evident in the different
implementation versions in Figure~\ref{fig:cf-versions}.  The page
lock level is finer than the size-class lock level, which exists in
all implementations.  In the presence of page locks, during compaction
the size-class lock is released and only the page locks of the source
and target page are locked.  As a result, other threads may perform
memory operations within the size-class that do not affect the source
and the target page.

Our managing of the global lists of empty pages and free abstract addresses is
inspired by the free list implementation used in~\cite{Hud2}.  Each of the two
lists is organized on two (public and private) levels.  Each thread owns one
private list (of free elements) which is only accessible to the owner thread.
Therefore the access to free elements in the private list needs no
synchronization mechanisms.  The public list is a list of lists of free
elements.  Its head contains a version number (used for synchronization between
threads) and a reference to the first element (sublist) in the list.  Both
fields in the list head are updated simultaneously using a double-word
compare-and-swap operation, hence the update is atomic.  Whenever the reference
to the first element changes, the version number increases, which prevents the
ABA problem~\cite{Herlihy08}.  If a thread needs a free element, then it first
accesses its private list.  If the private list is empty, then it accesses the
public list, in order to fetch the head of the public list of lists.  After
this, the newly fetched list becomes the private list of the thread.  There is
also a mechanism that returns elements from a private list to the public list,
which is invoked if the private list grows beyond a predefined bound.

There is a slight difference in the implementation of the public list
for the list of empty pages and for the list of free abstract
addresses.  In order to represent the public list in memory, we need
for each sublist a pointer to the next sublist.  In case of the list
of empty pages, we use the empty space of the first page of each
sublist to store such a pointer.  For the list of free abstract
addresses, an additional two-word data structure for storing the
pointers is needed.

\subsection{Concurrent Incremental CF}\label{subsec:conc-inc-cf}

For incremental compaction, each page-block stores an additional field
called compaction-block field.  The field has a size of 4B, which is
relatively small compared to the size of the page-blocks in
size-classes with large page-blocks (larger than 1KB), which are
typically subject to incremental compaction.  If a page-block becomes
a source/target of an incremental compaction operation, then its
compaction-block field stores a reference to its corresponding
target/source page-block, respectively.  Whether a page-block involved
in incremental compaction is a source or a target page-block is
determined by the status of its page and the status of the page of its
compaction block.

In addition, each abstract address contains a flag bit which signals
whether the object that the abstract address refers to is a target of
a canceled incremental compaction operation.  We have discussed
deallocation and compaction conflicts in Section~\ref{sec:inc-cf}.  In
the implementation, a deallocation conflict is detected if the
compaction-block field of the page-block under deallocation contains a
memory reference.  A compaction conflict is also recognized by a
memory reference in the compaction-block field of the source
page-block under compaction.  In case of a deallocation or a
compaction conflict, an ongoing compaction operation needs to be
canceled.  This is done by setting the flag bit in the abstract
address of the object that was deallocated and triggered the
compaction operation.  When the thread in charge of the canceled
compaction gets to execute again, it first checks the flag in the
abstract address and if the flag is set the thread terminates its
compaction operation and releases the abstract address.

\subsection{Local vs. Global Size-classes}\label{subsec:local-global}

An orthogonal optimization for concurrent CF which improves
scalability is using thread-local size-classes.  Every thread has a
private heap organized in private size-classes.  Each thread allocates
only in its private heap, but may deallocate shared objects in other
thread's heaps.  If the percentage of shared objects in the system is
low, this optimization leads to less conflicts, thus improving the
overall performance.

\section{Experiments}
\label{sec:exp}

We report on micro- and macrobenchmarks with non-concurrent non-incremental CF,
concurrent non-incremental CF, and microbenchmarks with concurrent incremental
CF.

\subsection{Hardware Setup}

The experiments with {\em{concurrent non-incremental}} CF ran on a server machine with
two quad-core 2GHz AMD Opteron processors and 16GB of memory. The experiments
with non-concurrent and non-incremental CF and concurrent incremental CF were
conducted on an XScale PXA 270 CPU with 600MHz and 128MB of memory.  The
operating system for both machines was Linux with real-time preemption patches
applied~\cite{RT}.  On the Opteron machine and the XScale machine the Linux
kernel version was 2.6.24 and 2.6.21, respectively.  In all experiments the
benchmark threads were executed with real-time priorities.

{We use two different processors since we are interested in
  evaluating the behavior of concurrent CF versions both on a
  multi-core server and on an embedded processor. The multi-core
  server has high computational power which removes the need of
  incremental compaction. These experiments are shown in
  Section~\ref{exp:cni}. To demonstrate the behavior of incremental CF
  in an embedded environment
  we use the XScale processor. These experiments are shown in
  Section~\ref{exp:ci}. Note that we compare the different concurrent versions of CF among themselves. See~\cite{USENIX08} for a comparison of the original CF with other allocators.}

\subsection{Concurrent Non-incremental CF}\label{exp:cni}

The microbenchmarks all run mutator threads that each allocate 2048
objects of random size, then deallocate the objects, and then start
over again.  The sizes of allocated objects correspond to the
distribution of object sizes allocated in a popular optimizer for
programmable logic arrays called Espresso used in several memory
allocator performance evaluations, e.g. in~\cite{Joh1}.  Each
microbenchmark runs for ten seconds performing more than one million
allocation/deallocation operations.

\begin{figure}
\begin{center}
\includegraphics[width=\textwidth]{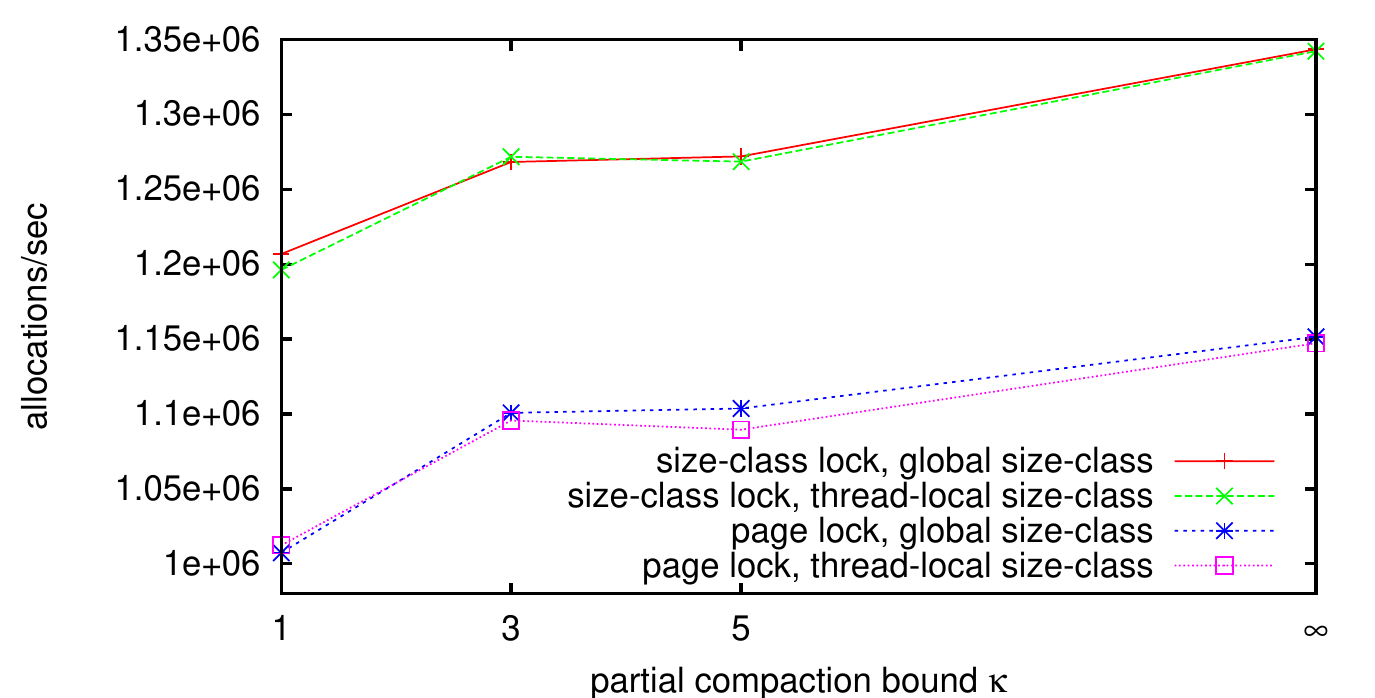}
\caption{Allocation throughput of a single thread with decreasing partial compaction}
\label{fig:exp-partial}
\end{center}
\end{figure}

Figure~\ref{fig:exp-partial} shows the impact of partial compaction on
the allocation throughput of a single thread.  Larger partial
compaction bounds $\kappa$ provide higher allocation throughput
because of less compaction activity.  Independently of $\kappa$, the
size-class lock configuration performs better then the page-lock
configuration since the latter needs locks for both the size-class
and the source and target pages.

\begin{figure*}
\begin{center}
\subfigure[full compaction\label{fig:throughput_k1}]{
\includegraphics[width=.6\textwidth]{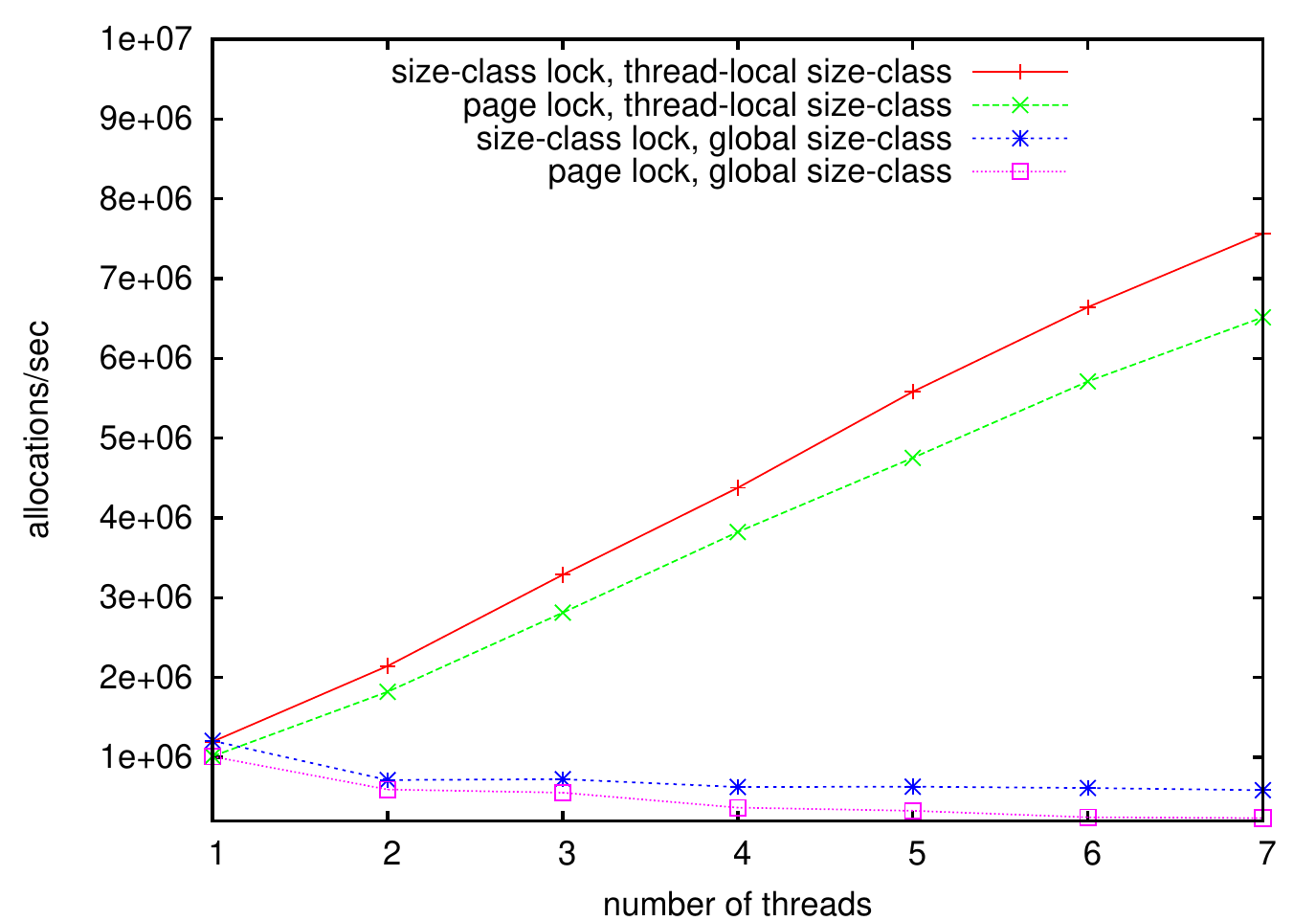} }
\subfigure[optimized, non-compacting\label{fig:throughput_k-1}]{
\includegraphics[width=.6\textwidth]{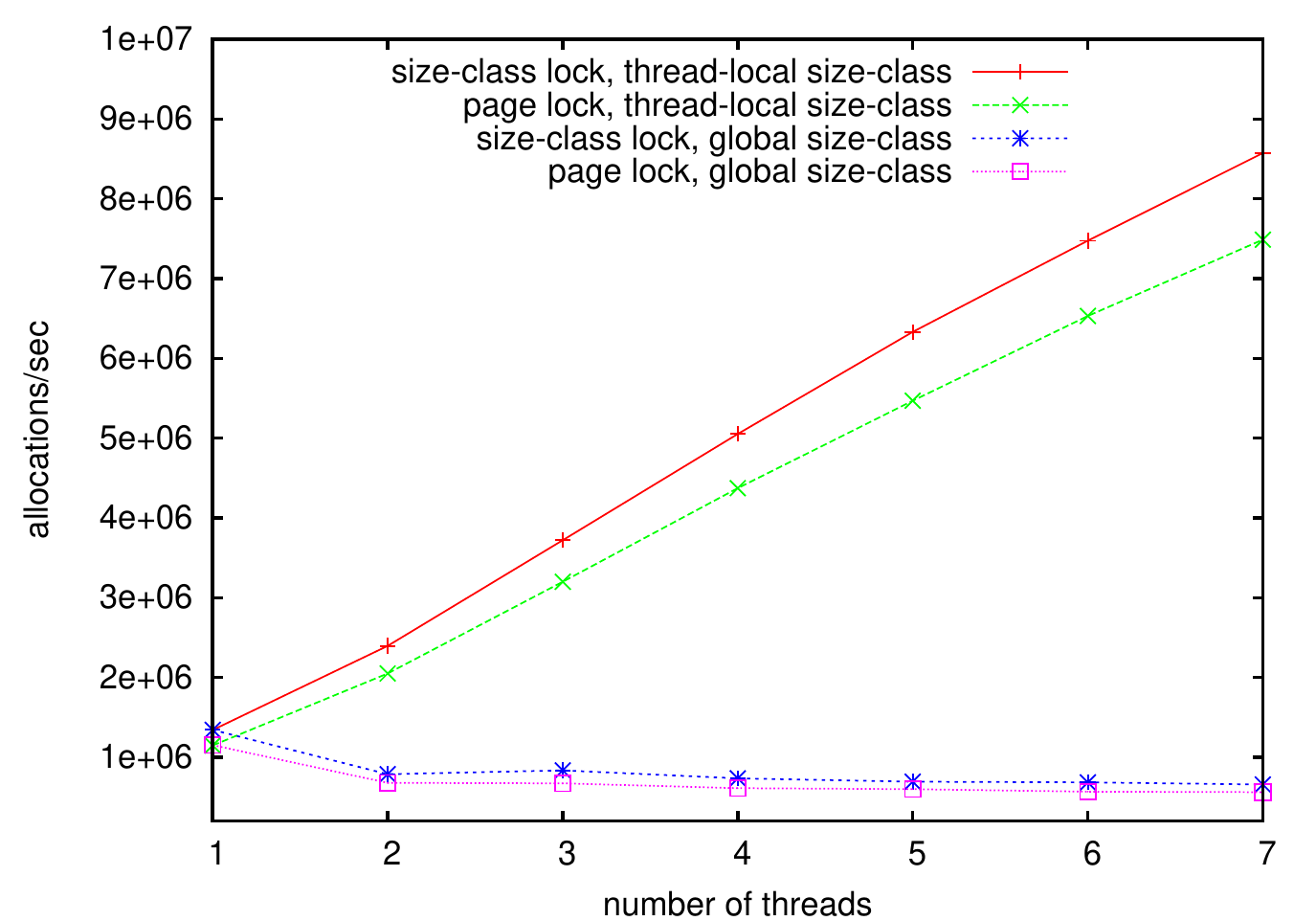}}
\subfigure[opt., non-comp. with sharing\label{fig:sharing}]{
\includegraphics[width=.6\textwidth]{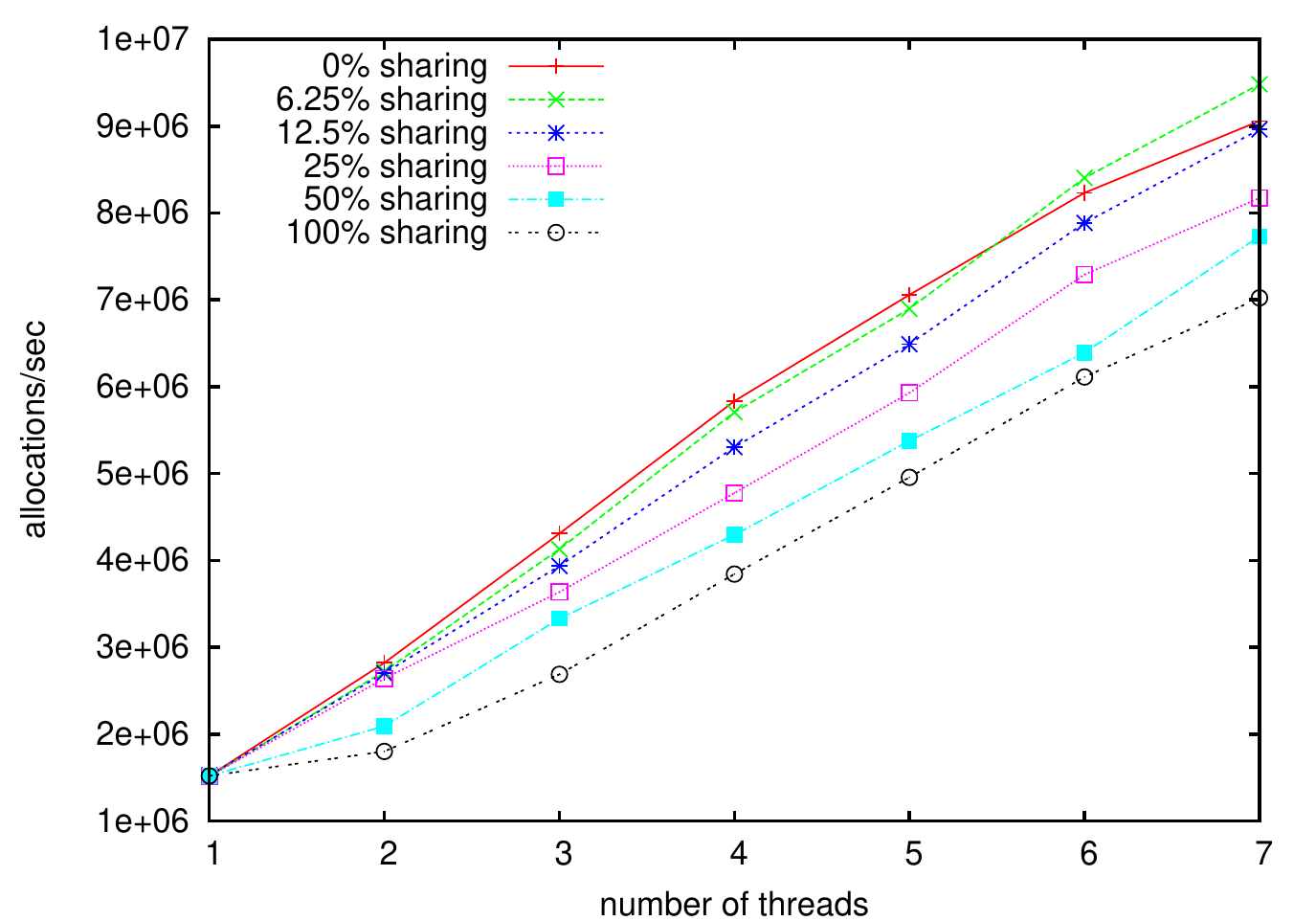}}
\caption{Allocation throughput with an increasing number of threads}
\label{fig:exp-scale}
\end{center}
\end{figure*}

Figure~\ref{fig:exp-scale} depicts the allocation throughput with an
increasing number of threads.  Up to seven threads run in parallel on
seven cores while the eighth core is used to minimize the influence of
collecting data on the performance data.  The performance of the fully
compacting and the optimized, non-compacting version of CF without
abstract addressing (in both cases with no sharing across the
thread-local CF instances) are shown in
Figures~\ref{fig:throughput_k1} and~\ref{fig:throughput_k-1},
respectively.  The thread-local size-class versions show linear
scalability in the number of threads whereas the global size-class
versions neither scale in the fully compacting nor in the
non-compacting configurations.  Again, the size-class lock
configurations result in better allocation throughput than the page
lock configurations.  Scalability only improves by a constant factor
with increasing partial compaction
(cf. Figures~\ref{fig:throughput_k1} versus~\ref{fig:throughput_k-1}).
Scalability of the thread-local size-class versions depends on the
degree of sharing across the thread-local CF instances.
Figure~\ref{fig:sharing} shows allocation throughput at varying
degrees of sharing: mutator threads allocate and deallocate 512
objects periodically according to the Espresso object size
distribution.  Each mutator frees its own just allocated objects and
objects previously allocated by other threads in a ratio that
determines the degree of sharing.

\begin{figure}
    \begin{center}
\includegraphics[width=\textwidth]
{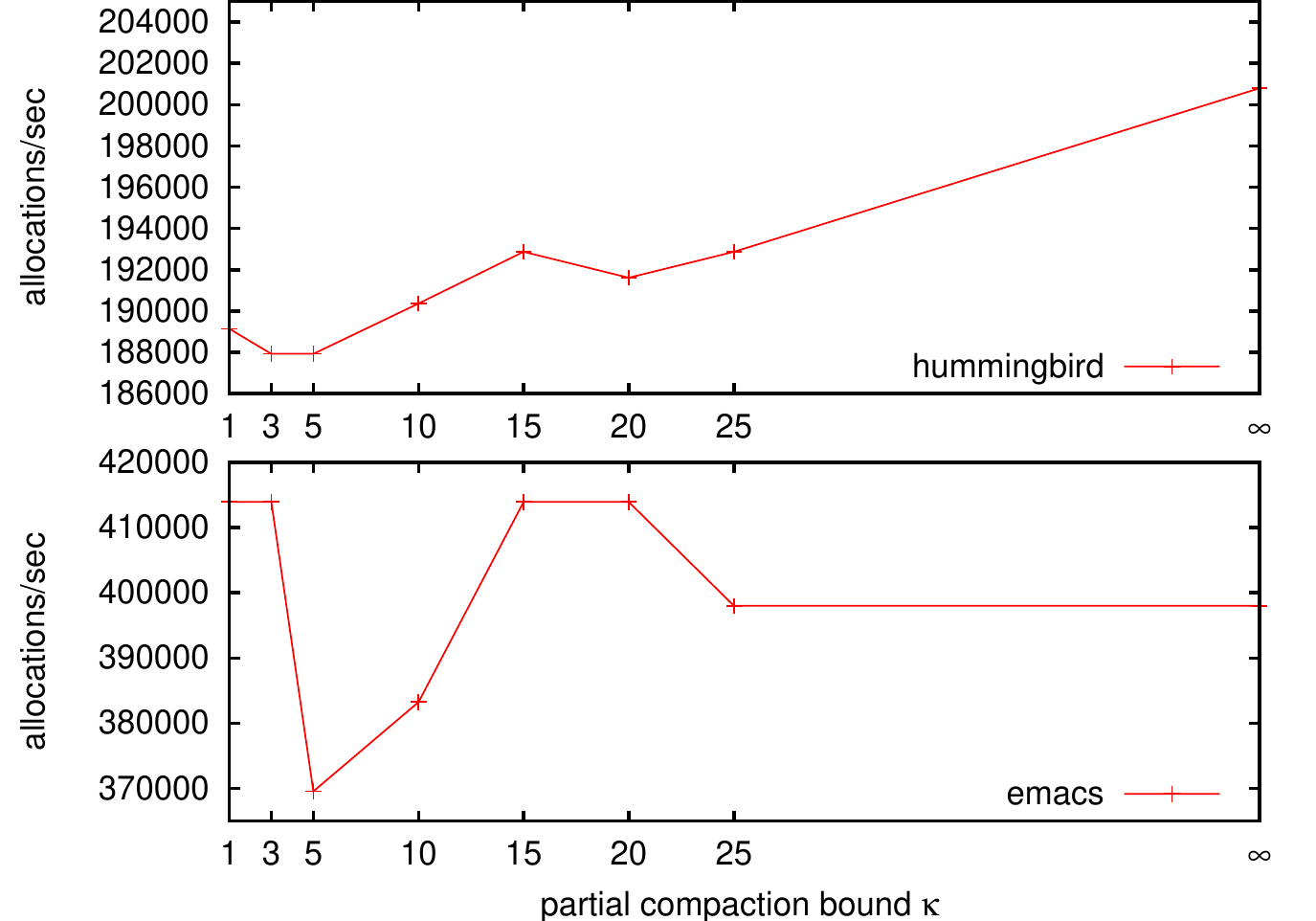}
        \caption{Allocation throughput for Hummingbird and Emacs}
        \label{exp:alloc-he}
    \end{center}
\end{figure}

The macrobenchmarks are based on Emacs and Hummingbird
allocation/deallocation traces~\cite{Boh1}.  In the Emacs trace about
51\% of the allocated objects are of size 40B, 15\% are of size 648B,
and 11\% are of size 104B.  The remaining objects of the trace are
also of small size.  In the Hummingbird trace about 25\% of the
allocated objects are of size 8B and 23\% are of size 32B.  The
remaining allocation requests vary from 16B to around 38.1MB (object
sizes greater than 16KB are ignored here).  Hummingbird's allocation
behavior is very different from the behavior of a typical mutator
where 99\% of the objects are of small and similar sizes~\cite{Joh1}.

Figure~\ref{exp:alloc-he} shows the allocation throughput of a single
thread running the Hummingbird and Emacs benchmarks.  Larger $\kappa$
values allow the Hummingbird benchmark to allocate more objects per
second.  In the Emacs benchmark the allocation throughput does not
improve for larger $\kappa$.

\begin{figure}
    \begin{center}
\includegraphics[width=\textwidth]
{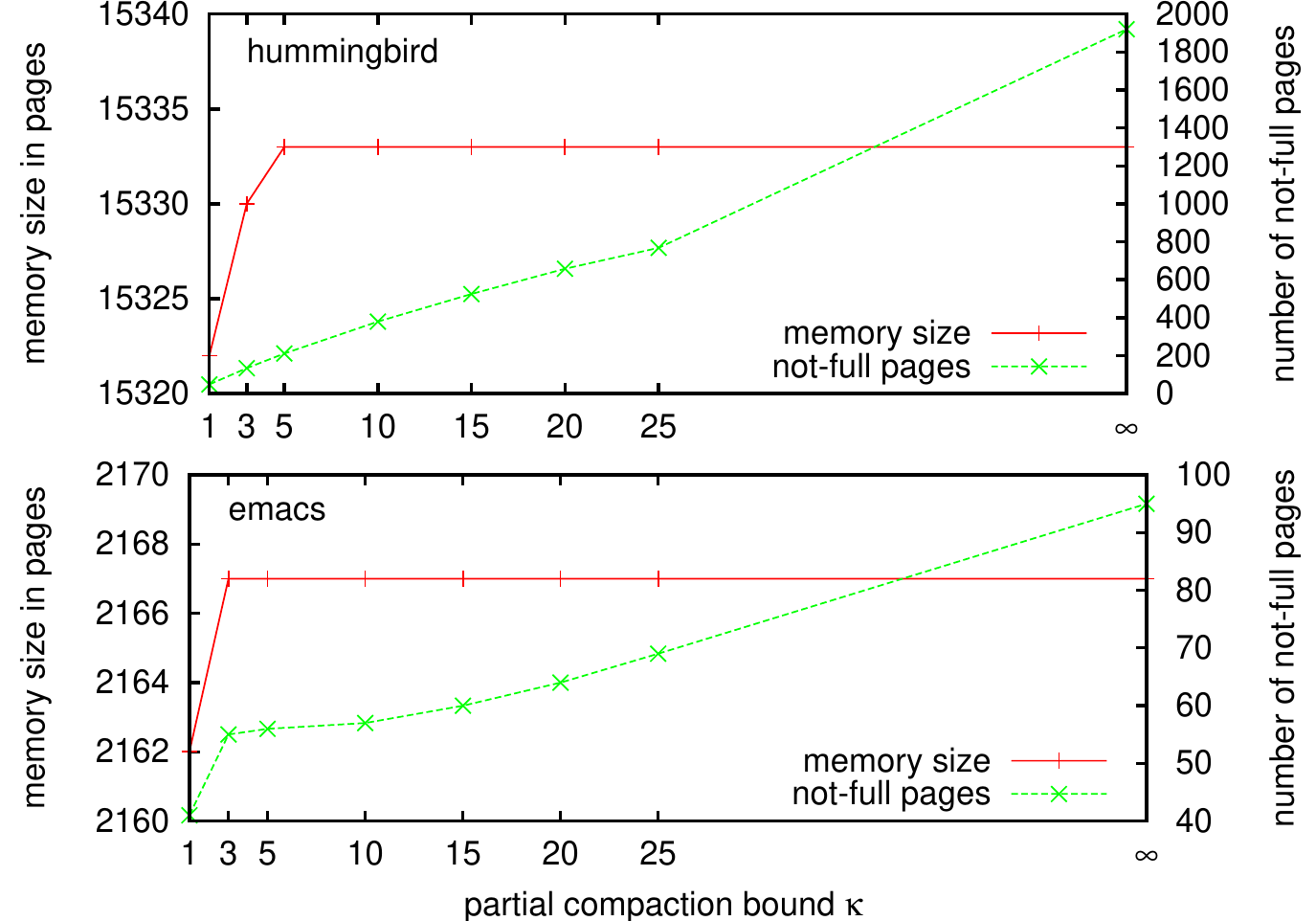}
        \caption{Memory usage and size-class fragmentation for Hummingbird and Emacs}
        \label{fig:exp-mem}
    \end{center}
\end{figure}

Figure~\ref{fig:exp-mem} shows the required memory size (in number of
used pages) and size-class fragmentation (in number of not-full pages)
during the execution of the Hummingbird and Emacs traces with
increasing~$\kappa$.  As expected, size-class fragmentation increases
with increasing $\kappa$, whereas the required memory size remains
constant for $\kappa \ge 5$ with the Hummingbird trace and $\kappa \ge
3$ with the Emacs trace since most not-full pages with smaller
page-block sizes tend to remain relatively full (in line with our
probabilistic claims of Section~\ref{sec:prob}).

\begin{table*}
\small
\centering
\begin{tabular}{|c|c|c|c|c|c|c|c|c|}
  \hline
  \multirow{3}{*}{} & \multicolumn{4}{|c|}{malloc (in clock ticks)} & \multicolumn{4}{|c|}{free (in clock ticks)}\\
  \hline
   & \multicolumn{2}{|c|}{TLSF} & \multicolumn{2}{|c|}{CF} & \multicolumn{2}{|c|}{TLSF} & \multicolumn{2}{|c|}{CF} \\
  \hline
  & avg & max & avg & max & avg & max & avg & max \\
  & time & time & time & time & time & time & time & time \\
  \hline
  \hline
  Emacs       & 228 & 93359 & 260 & 81662 & 153 & 71159 & 279 & 74798 \\
  \hline
  Hummingbird & 411 & 109079 & 529 & 98820 & 500 & 69192 & 574 & 79914 \\
  \hline
\end{tabular}
\caption{Performance: TLSF versus optimized, non-compacting CF (without abstract addressing)\label{tab:perf-tlsf-opt-CF}}
\end{table*}

Finally, Table~\ref{tab:perf-tlsf-opt-CF} shows the results of
macrobenchmarking TLSF~\cite{Mas1} and the optimized, non-compacting
version of CF without abstract addressing (configured to 16B, and
alternatively to 32B, for the smallest page-block size). The
temporal performance of malloc and free operations (in clock ticks
measured on the Opteron machine) for TLSF and non-compacting CF is
similar with TLSF slightly outperforming CF (except for malloc in the
worst case where CF is slightly better).


\subsection{Concurrent Incremental CF}\label{exp:ci}

The microbenchmark runs mutator threads allocating and deallocating objects
from 16B to 16KB randomly, where 90\% of the allocated objects are smaller than
64B~\cite{Johnstone98}.  The threads operate on global size classes.

\begin{figure*}
\begin{center}
\subfigure[allocation throughput\label{fig:throughput_incremental_iv_nv}]{
\includegraphics[width=0.6\textwidth]
{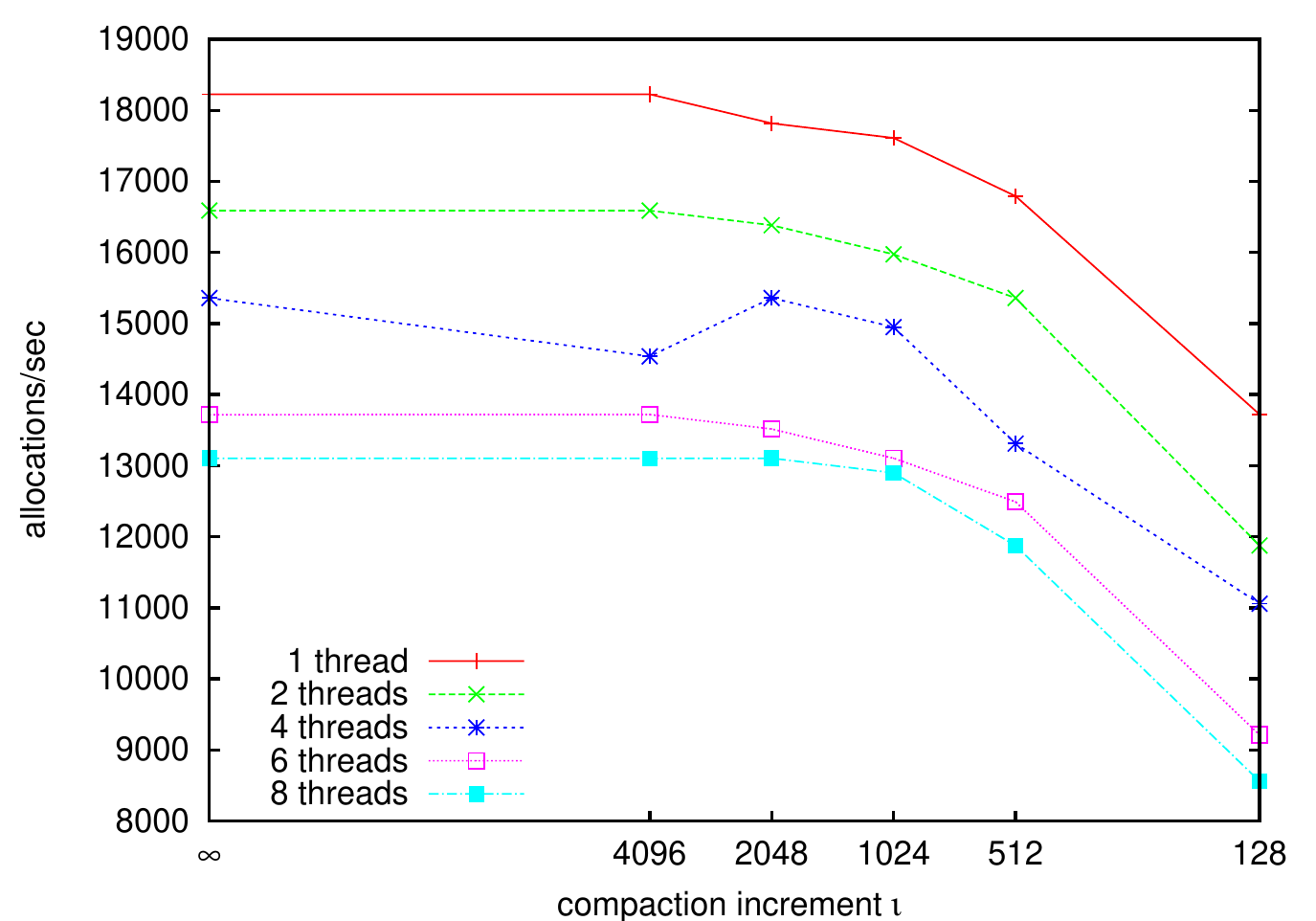}}
\subfigure[system latency with 8 threads\label{fig:latency_incremental_iv_nv}]{
\includegraphics[width=0.6\textwidth]
{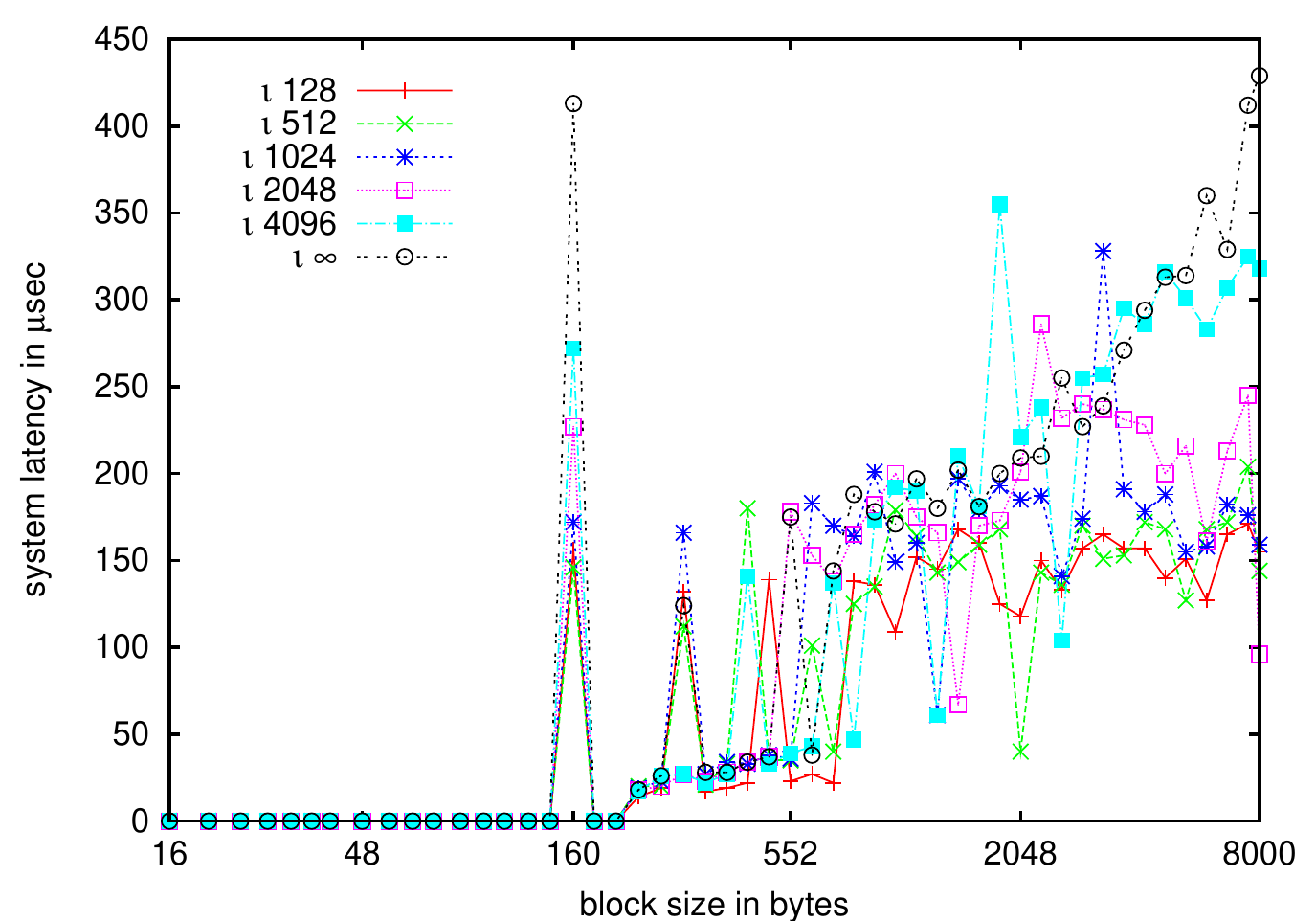}}
\subfigure[transient size-class fragmentation\label{fig:memory_incremental_iv_nv}]{
\includegraphics[width=0.6\textwidth]
{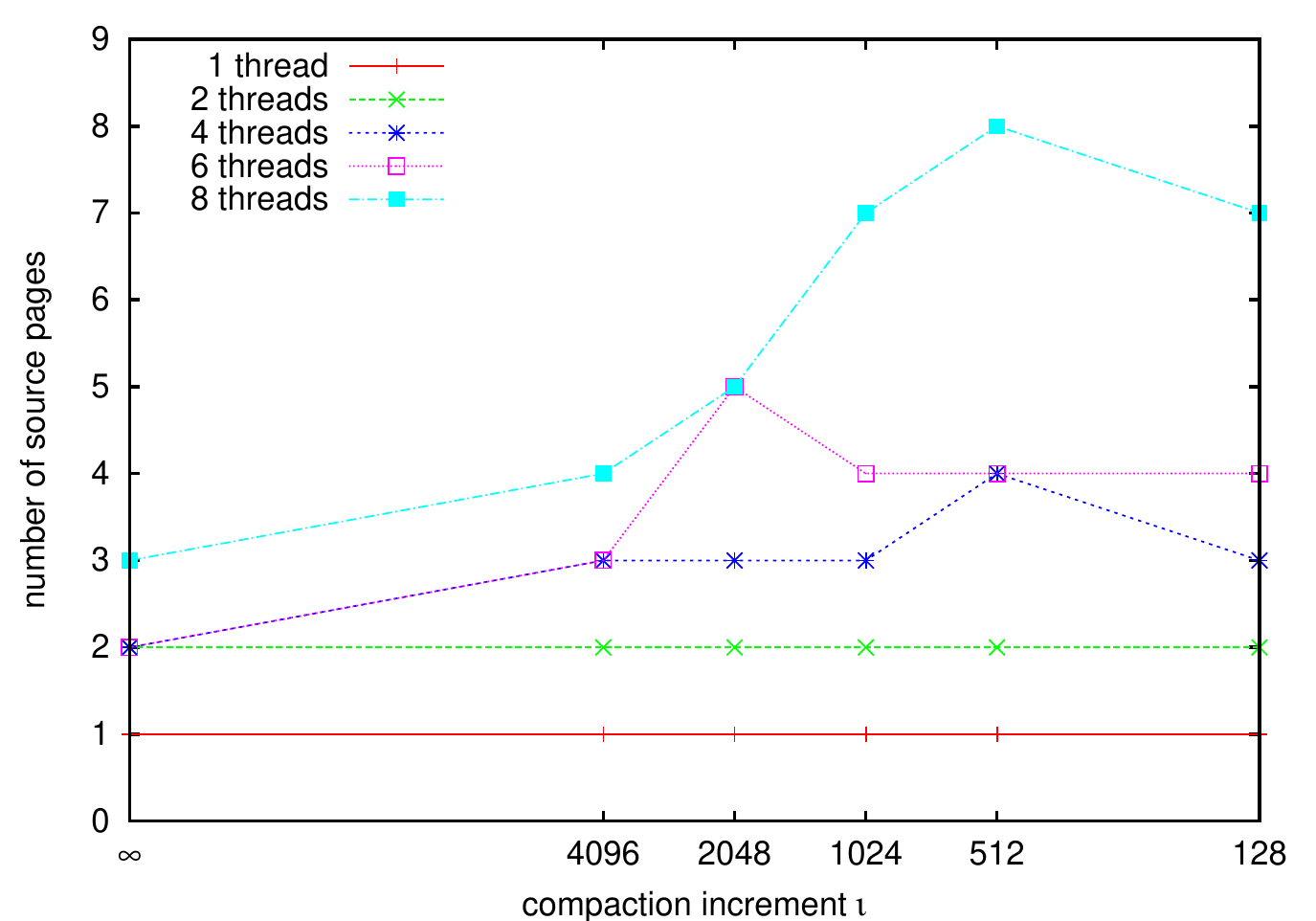}
}
\caption{Allocation throughput, system latency, and transient size-class fragmentation with decreasing compaction increments}
\label{fig:exp-inc}
\end{center}
\end{figure*}

Figure~\ref{fig:throughput_incremental_iv_nv} shows that the
allocation throughput decreases with decreasing compaction increments
$\iota$ since the incremental compaction overhead increases, due to an
increasing number of lock acquire/release operations, administrative
data updates, and memory copy interruptions.  System latency, shown in
Figure~\ref{fig:latency_incremental_iv_nv}, tends to decrease
measurably if page-block sizes larger than around 512B are involved,
with decreasing~$\iota$.  Here, we ran one mutator thread with higher
priority than seven other mutator threads, periodically yielding to
avoid starvation, and measured the maximum time the higher-priority
thread spent in the atomic portion of any incremental compaction
operation.  True system latency that includes the wait time for
locking was too noisy with the version of Linux we used.  Transient
size-class fragmentation, which is bounded by the number of threads,
generally increases slightly with increasing $\iota$ as shown in
Figure~\ref{fig:memory_incremental_iv_nv}.

\section{Conclusions}
\label{sec:conc}

Compact-fit is an explicit, dynamic heap management system that
allows, through the notion of partial and incremental compaction,
formally relating fragmentation, compaction, throughput, and latency
when managing contiguous blocks of memory.  We have studied this
relationship, formally and experimentally.  All versions of CF can be
made concurrent and scalable with partial compaction being only a
constant factor.  Scalability rather depends on the degree of sharing and
synchronization mechanisms, similar to other heap management
systems.

%

Incremental CF may open up a path to dynamic heap
management on memory-constrained systems running high-performance
applications that require tight temporal and spatial guarantees, although
further studies involving specialized operating system
infrastructure for embedded devices may be necessary there.

\bibliographystyle{amsplain}
\bibliography{paper}

\end{document}